\documentclass[pdflatex,sn-mathphys]{sn-jnl}

\jyear{2021}%

\theoremstyle{thmstyleone}%
\usepackage{graphicx,psfrag,color}
\usepackage{amsthm,amsmath,amsfonts,amssymb,float}
\usepackage[super]{cite}
\usepackage{fancyhdr}
\usepackage{amsmath,amsthm,amssymb}
\usepackage{mathrsfs}
\usepackage{indentfirst}
\usepackage{bbm}
\usepackage{subcaption}
\usepackage{natbib}
\usepackage{epsfig, epsfig, graphics, lscape, amsbsy, amstext}
\newtheorem{theorem}{THEOREM}

\newtheorem{lemma}{LEMMA}
\theoremstyle{thmstyletwo}%

\def\beqn{\begin{eqnarray}} 
\def\eeqn{\end{eqnarray}} 

 \def\be{\begin{equation}} 
 \def\ee{\end{equation}} 

\def\ge{\geq}

\makeatletter 
\renewcommand\@biblabel[1]{#1.} 
\makeatother

\begin{document}

\title[Article Title]{Long bet will lose: demystifying seemingly fair gambling via two-armed Futurity bandit}


\author[1]{\fnm{Zengjing} \sur{Chen}}\email{zjchen@sdu.edu.cn}
\author[2]{\fnm{Huaijin} \sur{Liang}}\email{huaijin@sdu.edu.cn}
\author[3]{\fnm{Wei} \sur{Wang}}\email{wangwei\_0115@outlook.com}
\author[4]{\fnm{Xiaodong} \sur{Yan}}\email{yanxiaodong@sdu.edu.cn}
\equalcont{These authors contributed equally to this work.}

\affil[1]{\orgdiv{Zhongtai Securities Institute for Financial Studies}, \orgname{Shandong University}, \orgaddress{\street{Street}, \city{Jinan}, \postcode{250100}, \state{Shandong}, \country{China}}}
\affil[1]{\orgdiv{National Center for Applied Mathematics}, \orgname{Shandong},  \country{China}}



\abstract{No matter how much some gamblers occasionally win, as long as they continue to gamble, sooner or later they will loss more to the casino, which is the so-called ``long bet will lose''. Our results demonstrate the counter-intuitive phenomena that
gamblers involve ``long bet will lose'' but casinos always advertise their unprofitable circumstances. Here we expose the law of inevitability behind ``long bet will lose'' by theoretically and experimentally demystifying the profitable mystery behind casinos under two-armed antique Mills Futurity slot machine\cite{Ethier2010}.
The main results straightforwardly
 elucidate that all casino projects are seemingly a fair gamble but essentially unfair, i.e., the casino's win rate is greater than 50\%. We anticipate our assay to be a starting point for studying the fairness of more
sophisticated multi-armed Futurity bandits based on mathematical tool. In application, fairness study of the Futurity bandits not only exposes the fraud of casinos for gamblers but also discloses the discount marketing, bundled sales or other induced consumption tactics.
}

\keywords{Fairness, Multi-armed bandit, Futurity slot machine}



\maketitle
\section*{Introduction}\label{sec1}

The history of human gambling is estimated to be about the same as that of civilization. Tracing back to the history of human gambling, people had the habit of ``taking chances'' as early as the late Paleolithic Age, such as divination was widely used to judge the bad and the good, including major wars, through divination
in prehistoric civilization of China. The appearance of casinos has boosted the prosperity of the gambling industry, which has been enduring.
Various forms of gambling have been derived, such as horse racing and lottery in Hong Kong, baccarat and dice in Macao, baccarat, slot machine, roulette, blackjack, etc., and the Philippine government supports and encourages the development of the gambling industry because they found that the gambling industry stimulates domestic economic development under the situation of global economic downturn, which essentially elucidates the profitability of gambling industry but a secret that the casino never said, no matter how they are changed, causing a situation of the so-called  ``long bet will lose'' for the players.

Gambling activity attracts amounts of players by the fairness illusion that the casino is unprofitable, then their enthusiasm is enlightened by the so-called honest advertisement, even they indulge in fantasy of earning more money. The rigged casino appears
to be especially absorbing to players with gambling-related pathology\cite{Dixonmike} who becomes completely immersed in the playing, then depression symptomatology, gambling expectancy and dark flow ratings are also emerging\cite{Dixon}. 
Such honest advertisement about the fairness can be attributed to the casino loyalty programmes \cite{Hollingshead} offering the same rewards to gamblers
for their equivalent input. 
The aim of a loyalty programme is to increase 
attitudinal and behavioural loyalty. Attitudinal loyalty 
refers to the degree that people trust and are satisfied 
with the casino. It also involves a sense of 
identification with the casino brand. Behavioural 
loyalty refers to people's actual behaviours that show 
their loyalty. For example, visiting the casino again and again to 
gamble. However, all seemingly fair casino projects are essentially unfair, because it has never been heard that the casinos are defeated by players, until the occurrence of Kelly formula\cite{Kelly} with long reputation in ``Las Vegas'' and ``Wall Street'' delivers the optimal proportion that should be wagered in each period\cite{Thorp} in a repeated gamble of  Blackjack (21 points) or repeated investment with more than 50\% winning rate.

Enlightened by the boom of gambling industry nevertheless the attitudinal loyalty of the players on the casino, this article explores the potentially profound mystery with a multi-armed Futurity bandit mathematically. Because 
multi-armed bandit (MAB)\cite{feller,Sut2018}, as a popular entertainment tool, has been meticulously designed by the casinos for looking seemingly fair and attracting the gamblers\cite{Ethier2010}  .
MAB has also been widely studied theoretically to analyze amounts of complex decision problems\cite{chenepstein,ACF,Chen-Feng-Zhang} in science, society, economy, management and other areas and plays a central role in reinforcement learning\cite{Morris,Narisawa,Aga1995,Collins} . 
In detail, this work introduces the two-armed Futurity bandit to demystify that
such all-encompassing absorption of the gamblers on the casino
can be attributed to the seemingly fair one-armed slot machine,
because it can be unprofitable for casino by the futurity reward design, such as the futurity slot machine, when the current number of consecutive failures reaches $J$, all $J$ coins will be
returned at this time.
However, two-armed slot machines break the fairness and shows Parrondo's paradox phenomenon\cite{Parrondo} that the casino becomes profitable when players alternate arms in certain random or nonrandom ways as well as being
advertised that two arms are fair.


People just learned from experience that ``long bet will lose'', but
 the law of inevitability behind ``long bet will lose'' is determined by the ``law of large numbers'' in probability theory and the mystery behind casinos will be unknown without the probability.  
This article adopts probability tool the law of large numbers and
two-armed antique Mills Futurity slot machine, designed by the Chicago Mills Novelty Company\cite{Ethier2010}. The Futurity slot machine
applies fair compensation after consecutive failure to reveal the seemingly fair fraud of gambling, which alludes to ``\textit{fairness illusion}''. The compensation rule satisfies that two coins will be returned to the player when the consecutive failures
reach two correspondingly, and the fraud can be stated by the followings. If we play one-armed Futurity bandit, the smart casinos always honestly advertise that the deviced slot machines
are unprofitable through advocating the fairness, which implies that the casino isn't capable of making money in the long run through
using one-armed Futurity bandit for gambling. This trick designed on one-armed Futurity bandit increases the reputation of casino in terms of the fairness, whilst  artifacts occur for two-armed Futurity bandit by alternatively playing the left and right arms, which demonstrates that the casino can be always profitable under the same rule that if consecutive failures occur two times, the player will obtain the common return with two coins. 
This results are corresponds to the conjecture proposed by S. N. Ethier and Jiyeon Lee's\cite{Ethier2010} showing that this two-armed Futurity bandit obeys Parrondo’s paradox when
it repeatedly executes any non-random mixed strategy $D$ and $J = 2$. Specifically, a non-random strategy decided by the player before playing including two arms, such as $D = ABB$, and then repeats the game according to
the strategy $D$, returning two coins for every two consecutive failures during the
game. This article mathematically proves this conjecture and uses experiment to verify the theoretical results.

The main contributions of this article include the followings:
\begin{itemize}
    \item This the first development of nonrandom two-armed Futurity bandit to reveal the profitable law of seemingly fair   design of gambling slot machine, enlightening inspiration source of the
gambling boom and the miserable lot ``long bet will loss'' for the players.
    \item This work proposes a novel probability framework for a nonrandom sequences of arm-playing rule and provides the explicit expressions of profitable return for casinos, which elucidates the changing law of the return along with the considered parameters.
\item This assay would be a starting point for more 
sophisticated practical studies by using two-armed Futurity bandit, such as disclosing the discount marketing, bundled sales or other induced consumption tactics.
\end{itemize}

\section*{Results}\label{sec2}

{\bf Model.} As a casino project, the Futurity slot machine was designed by the Chicago Mills Novelty Company. A player consumes 1 coin per coup and there are two screens on the slot machine. The pointer in one screen records the current number of consecutive failures. When the number reaches 10 (it can also be other values set by the casino), all 10 coins will be returned at this time. This payoff is called the Futurity award. Another screen will display the current mode. Its internal structure is a periodic cam with several fixed modes. Different modes will have different winning conditions and rewards for the current game. Each time the player plays a game, the cam rotates to the next pattern. The player chooses to pull the arm once before playing the game, and each arm has a separate mode cam that is different from each other, which means that each arm has an independent payoff distribution. For the Futurity award, regardless of the order in which the player plays the slot machine, the current number of consecutive failures will be recorded, and when the pointer reaches the value $J(J\geq 2)$ set by the casino, the $J$ coins will be returned to the player. The casino owner honestly advertises that all arms on his multiple-armed machine are ``fair'' in the sense that either of the arms has a 50 $\%$ chance of making the player profitable. The gambler is allowed to play either arm on every play in some deterministic order or at random. An important question would be whether such a casino still obeys the rule that ``long bet will lose'', how much profit can casino owner could win?

For simplify, we consider a simple two-armed Futurity bandit, where the two arms denoted as $A$ and $B$ and each arm has a different i.i.d payoff sequence. For the convenience of analysis, we can regard it as a Bernoulli distribution, that is, the probability of winning a game is $p_A$ (resp. $p_B$), and $0<p_A, p_B<1$. The player needs to pay 1 coin to the casino at each game and shakes the two arms alternatively under a pre-determined playing rule.  The casino also brings futurity reward to the player who suffers consecutive failure in gambling, that is, if the player fails in $J$ consecutive games, the casino will return all the coins spent in the $J$ games.  The casino usually advertise to design $J=2$, which is the considered case in this work and the most attractive for the gamblers.
Before the game starts, the player needs to choose a pre-formulated non-random mixed strategy $D$, where $D$ contains at least 1 $A$ and 1 $B$, such as $D=ABB$ meaning that the players will proceed to pull the arms $A$, $B$ and $B$ repeatedly and indefinitely.
This work consider a ``fairness'' design for one-armed Futurity bandit by adjusting the payoff distribution of each arm, where the 
 reward is assumed to be $(3-2p)/(2-p)$ under the wining probability $p=p_A$ and $p_B$ for Arms $A$ and $B$ respectively. 
 Once the gambler plays the game according to the above rules, it seems that the gambler is fairly playing the game with no loss, but the casino will definitely make a profit in the long run, which is demystified in the following theorem.

Subtly mathematical deduction can conclude that any non-random mixed strategy $D$ can be always arranged by the following same asymptotic form $D(\boldsymbol{a}(h,r,s))$, that is
\begin{equation}\label{eq1}
D(\boldsymbol{a}(h,r,s))=\underbrace{A...A}_{r_1}\underbrace{B...B}_{s_1}...\underbrace{A...A}_{r_k}\underbrace{B...B}_{s_k}...\underbrace{A...A}_{r_h}\underbrace{B...B}_{s_h}
\end{equation}
where $r_k>0$, $s_k>0$, $\sum\limits_{k=1}^{h}r_k=r$, $\sum\limits_{k=1}^{h}s_k=s$ and the vector $\boldsymbol{a}(h,r,s)=(a_1,a_2,...,a_{2h})=(r_1,s_1,...,r_k,s_k,...,r_h,s_h)$, which is used to describe the structure of any strategy $D$.
In order to make our results more concise, we define the function $b_i$ of the vector $\boldsymbol{a}$ for $1\le i\le 4h$ as follows:
\begin{itemize}
\item ${b_{2j-1}}=(-1)^{a_{2j-1}}(1-p_A)^{a_{2j-1}}$, ${b_{2j}}=(-1)^{a_{2j}}(1-p_B)^{a_{2j}}$ for $1\le j\le h$.\\
\item ${b_i}={b_{i-2h}}$  for $2h+1 \le i \le 4h$.\\
\end{itemize}


\begin{theorem}\label{th1}
The casino's asymptotic profit expectation $R$ is $2QS$, where
$$Q:=Q(D(\boldsymbol{a}(h,r,s)))=h+\sum\limits_{m=1}^{2h}\sum\limits_{j=1}^{2h-1}(-1)^j\prod\limits_{i=m}^{m+j-1}{b_i}+h\prod\limits_{i=1}^{2h}{b_i}$$
$$S:=S(r,s,p_A,p_B)=\frac{(p_A-p_B)^2(1+(-1)^{r+s}(1-p_A)^r(1-p_B)^s)}{(r+s)(2-p_A)^2(2-p_B)^2(1-(1-p_A)^{2r}(1-p_B)^{2s})}$$

\end{theorem}
Theoretical results elucidate that the casino would be always profitable if $p_A\neq p_B$, and the asymptotic profit expectation $R=0$ if and only if $p_A= p_B$, which also implies that the discrepancy of the two arms in terms of the distinct winning probability favors the casino.
In detail, the expression of casino asymptotic profit expectation $R$ consists of three parts, where the first part is the number 2, representing the settlement rule $J=2$ of the Futurity bandit reward. The second part function $Q$ denotes the playing rule across the two arms in the internal structure of strategy $D$. The last part function $S$ characterizes the profitable change along with the considered parameters 
$p_A, p_B$ and the considered playing number $r,s$.

Fig. \ref{MC3D} shows the three-dimensional surface of the casino's payoff along with the winning probability $p_{A}$ and $p_{B}$ of Arms $A$ and $B$ respectively under four different but representative nonrandom strategy $D$. The four sub-images shows that different nonrandom strategies generate distinct profit modes with commonly positive profit of casino. Fig. \ref{MC3D}(a) implies that alternatively playing the two arms can be capable of generating more profits for the casino. Equally playing the two arms guarantees the symmetric form of the payoffs (see Fig.s \ref{MC3D}(a) and (b)).

\begin{figure}
\centering 
\subfloat[$p_{\gamma}=0.1$]{\includegraphics[width=5cm,height=5cm]{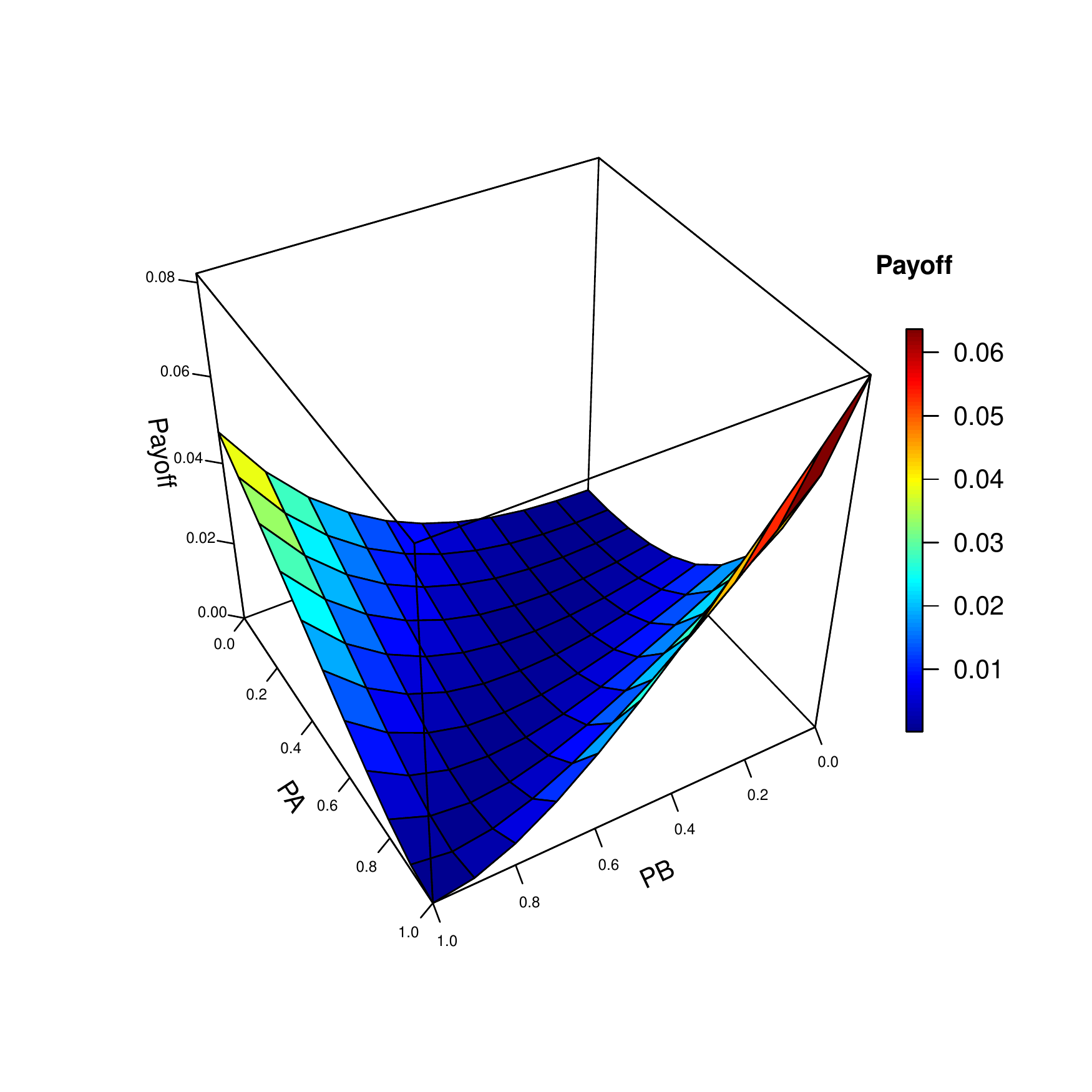}\label{subfig:1}}\hspace{5pt} \subfloat[$p_{\gamma}=0.3$]{\includegraphics[width=5cm,height=5cm]{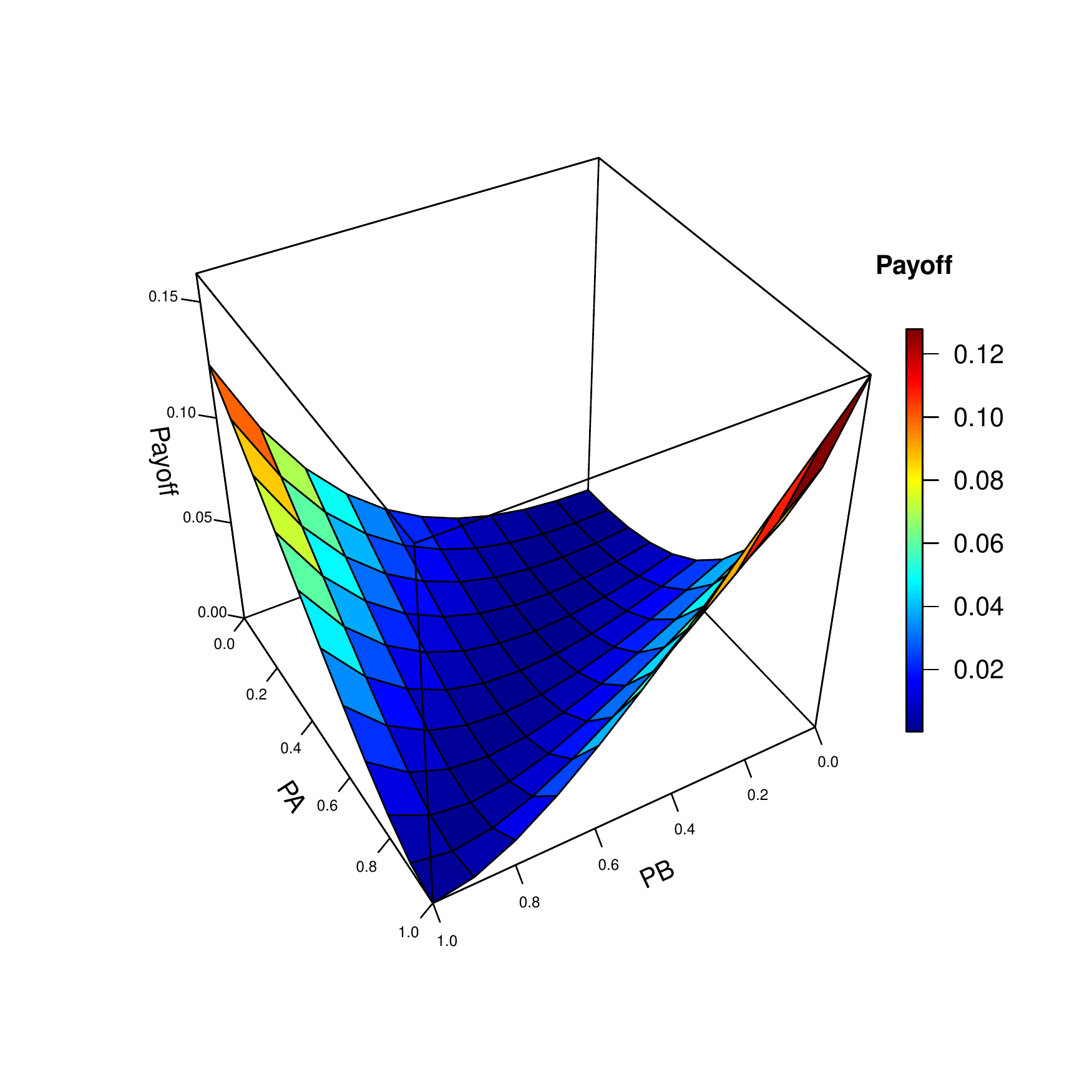}\label{subfig:2}}\\
\subfloat[$p_{\gamma}=0.5$]{\includegraphics[width=5cm,height=5cm]{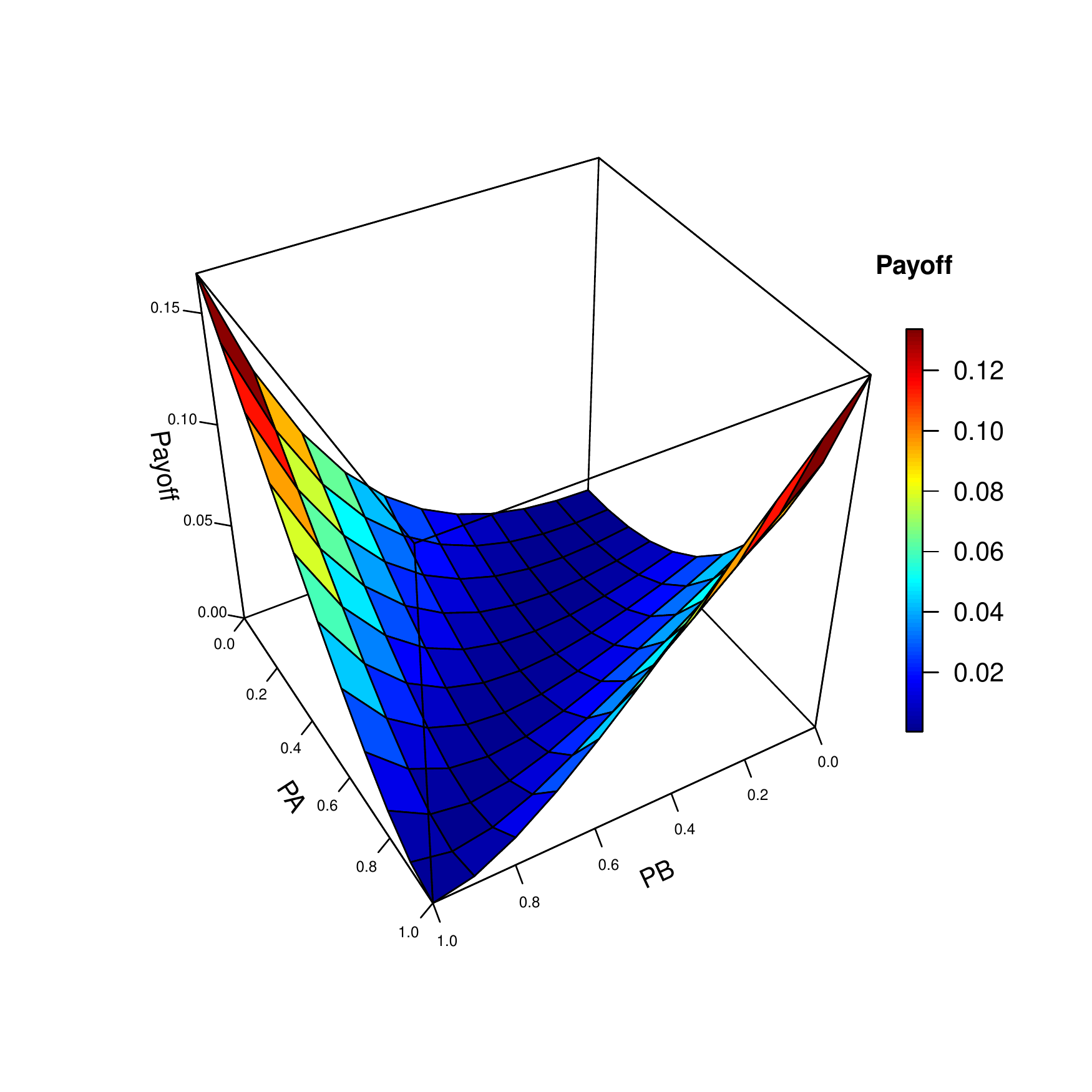}\label{subfig:3}}\hspace{5pt} \subfloat[$p_{\gamma}=0.7$]{\includegraphics[width=5cm,height=5cm]{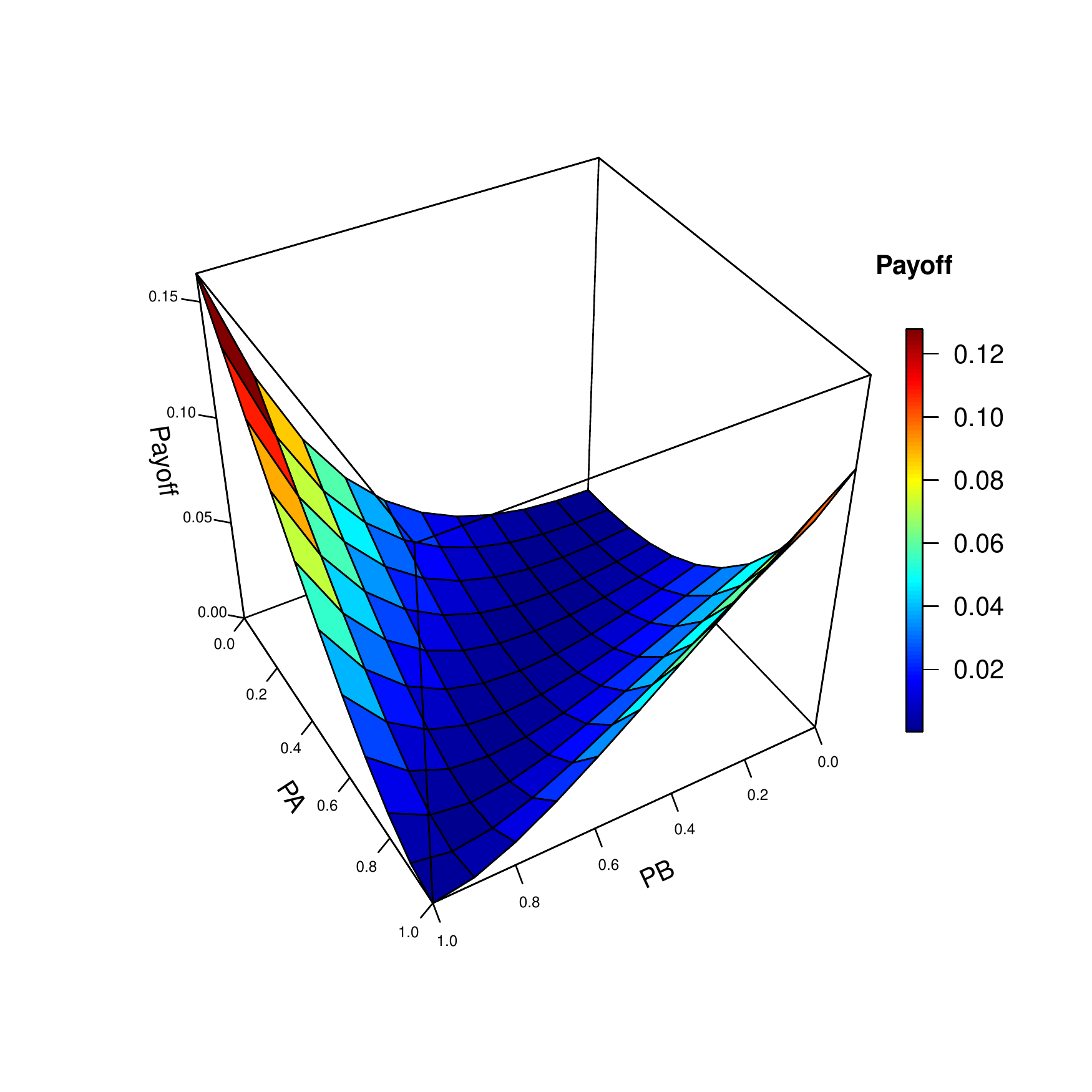}\label{subfig:4}}
\subfloat[$p_{\gamma}=0.9$]{\includegraphics[width=5cm,height=5cm]{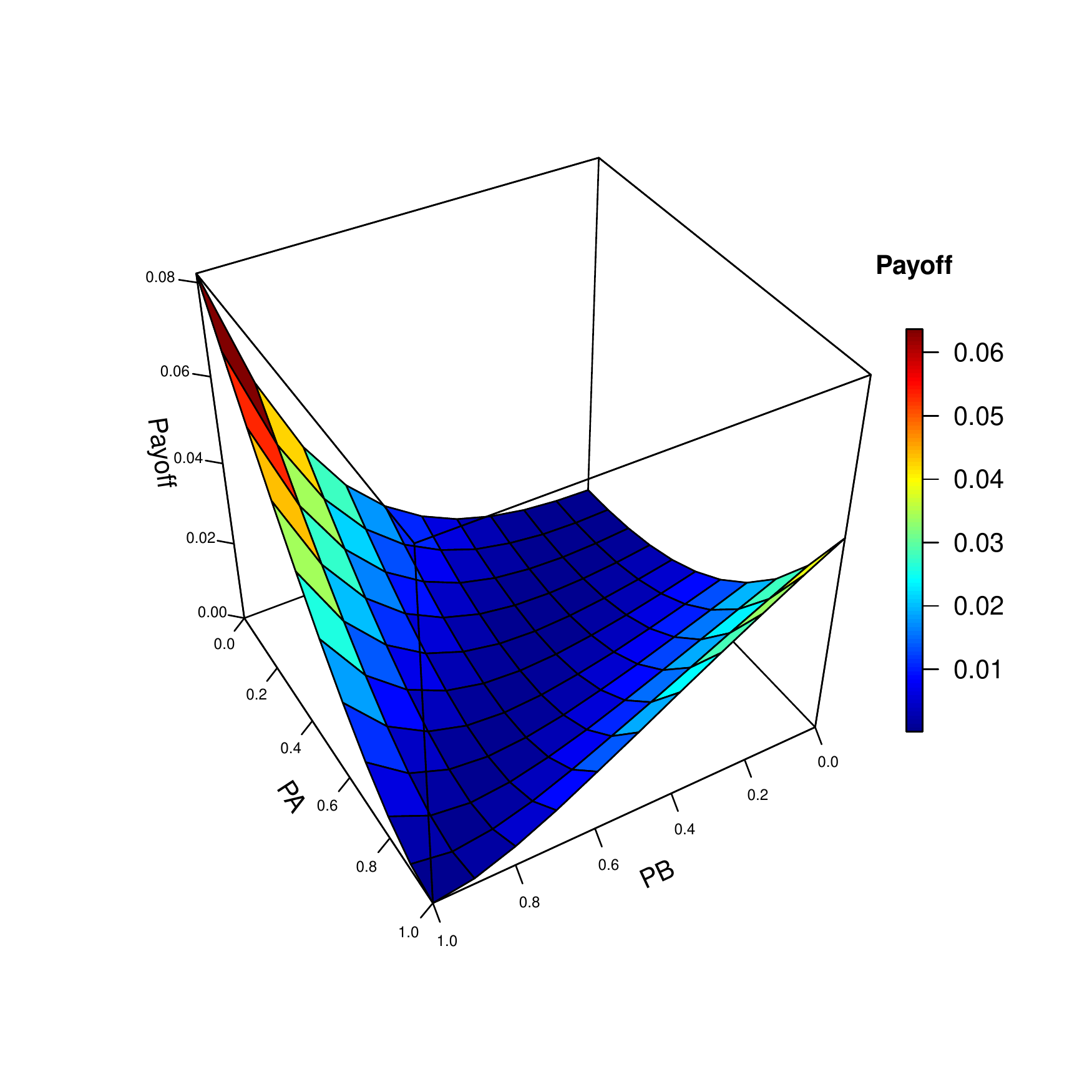}\label{subfig:5}}
\caption{\label{3DplotRandom}The sample mean payoff of casino under the random strategy with probability $p_{\gamma}$ selecting Arm $A$. }
\end{figure}

If the gambler plays the two-armed Futurity bandit according to a random strategy $C$ with the probability $p_\gamma$ playing Arm $A$ and correspondingly playing the Arm B with probability $1-p_\gamma$.
The asymptotic profit expectation $R_C$ of the casino has been concluded \cite{Ethier2010} is 
$$R_C=f(p_\gamma (1-p_A)+(1-p_\gamma)(1-p_B))-p_\gamma f(1-p_A)-(1-p_\gamma)f(1-p_B),$$
 where $f(z)=\frac{2z^2}{1+z}$. Since $f(z)$ is a convex function, the casino will be profitable in the long run with $R_C\ge 0$ and $R_C=0$ if and only if $p_A=p_B$. 
Fig. \ref{3DplotRandom} shows the payoff performance under random strategy with probability $p_{\gamma}=\{0.1,0.3,0.5,0.7,0.9\}$ selecting Arm $A$. All the sub-figures display non-negative payoffs under any combination of $p_A$ and $p_B$. $p_{\gamma}=0.5$ meaning equal numbers of playing the two arms generates a asymmetric images of the payoffs, which implies the same results corresponding to Fig.s \ref{MC3D}(a) and (b) in terms of the symmetry.

\section*{Experiment}\label{sec3}

\subsection*{The simulated proof of the theoretical results}\label{sec31}
This section designs
Monte Carlo simulations to verify the theoretical results under four cases corresponding to the non-random mixed strategies $D=AB$, $D=AABB$, $D=AAABB$, $D=AAAABBBBAAAAAABBB$ with the number of playing $M=100000$ and the sample mean of profit for casino equals $(M-W-J*C)/M$, where $W$ represents the total number of win for the gambler and  $C$ denotes the count of  consecutive $J$ failures, where $J=2$. The 10000 replications are conducted in each simulations.
  
\begin{figure}
\includegraphics[width=10cm,height=7cm]{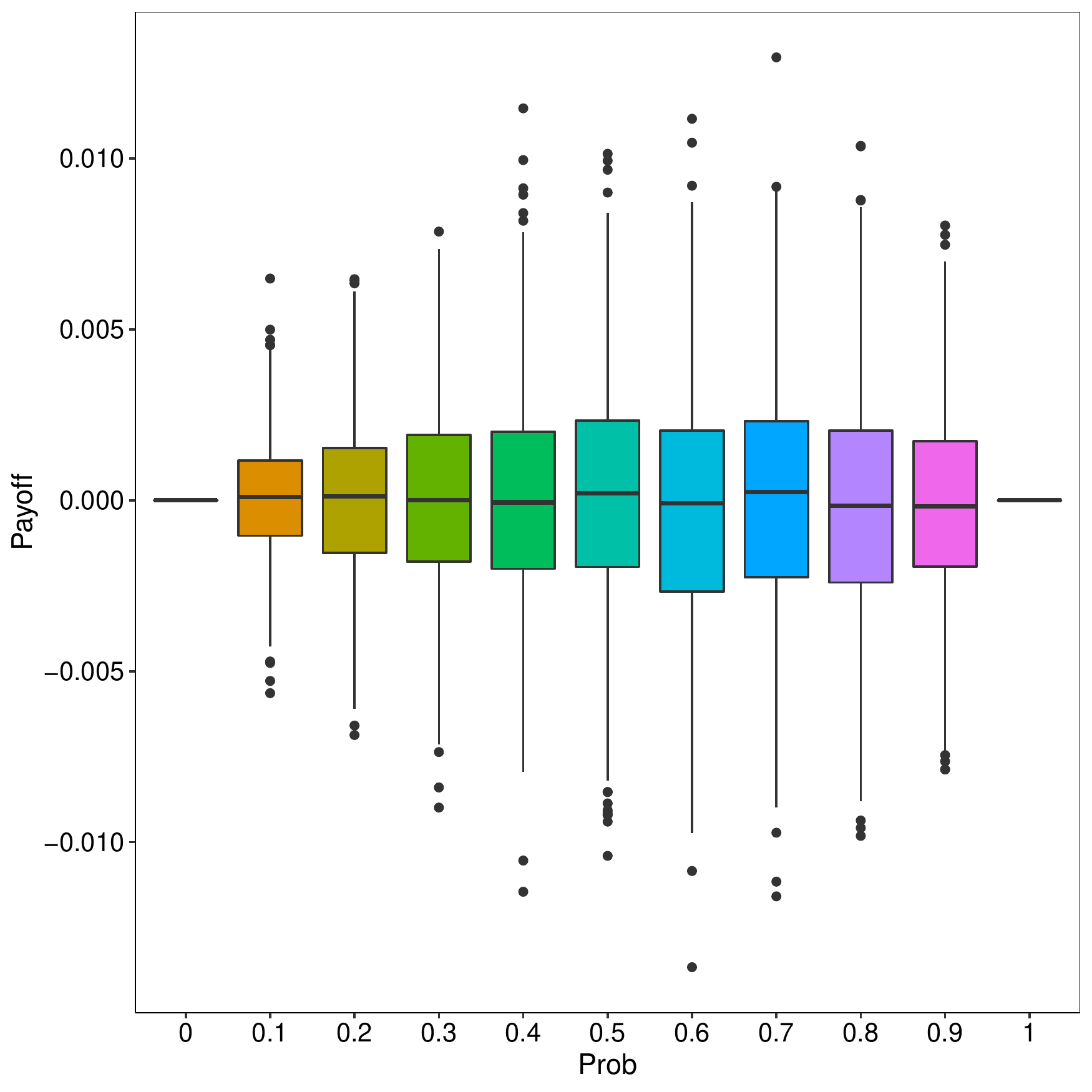}
\caption{\label{box}The payoff of single arm across probability $[0,1]$.}
\end{figure}

 As stated above, the casino can adjust the probability distribution of the two arms so that it can not only ensure the fairness of the two arms but also adjust its own profit. In the first simulation of single arm slot,  Fig. \ref{box} shows that, the payoff of player or casino always centres around zero across different probability in interval [0,1]  for the single arm, representing the fairness for each arm. In particular, the payoff of gambler or casino is zero without uncertainty, while the probability is 0 or 1. 
 
\begin{figure}[th]
\centering \subfloat[$D=AB$]{\includegraphics[width=5cm,height=5cm]{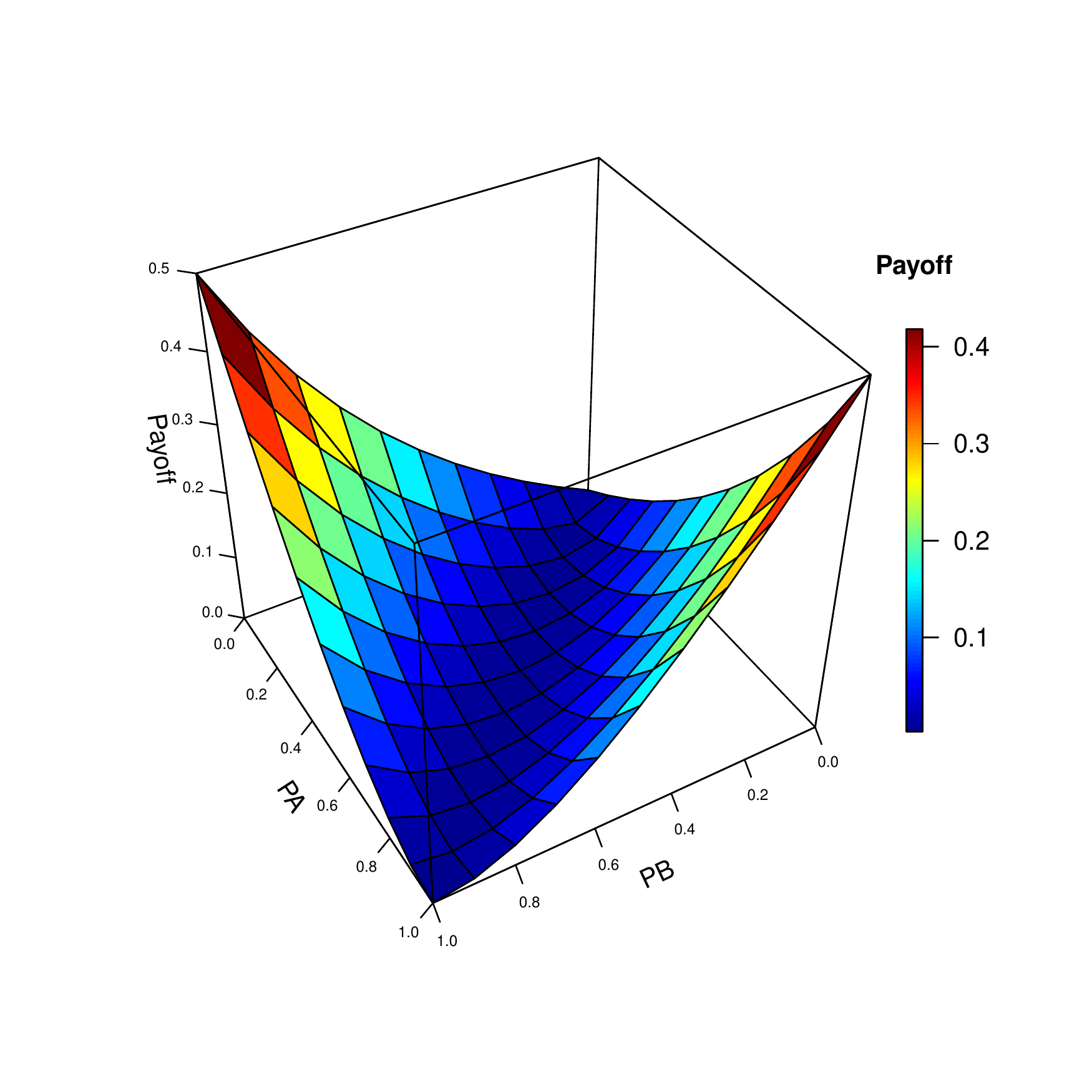}\label{subfig:1-1-1}
}\hspace{20pt} \subfloat[$D=AABB$]{\includegraphics[width=5cm,height=5cm]{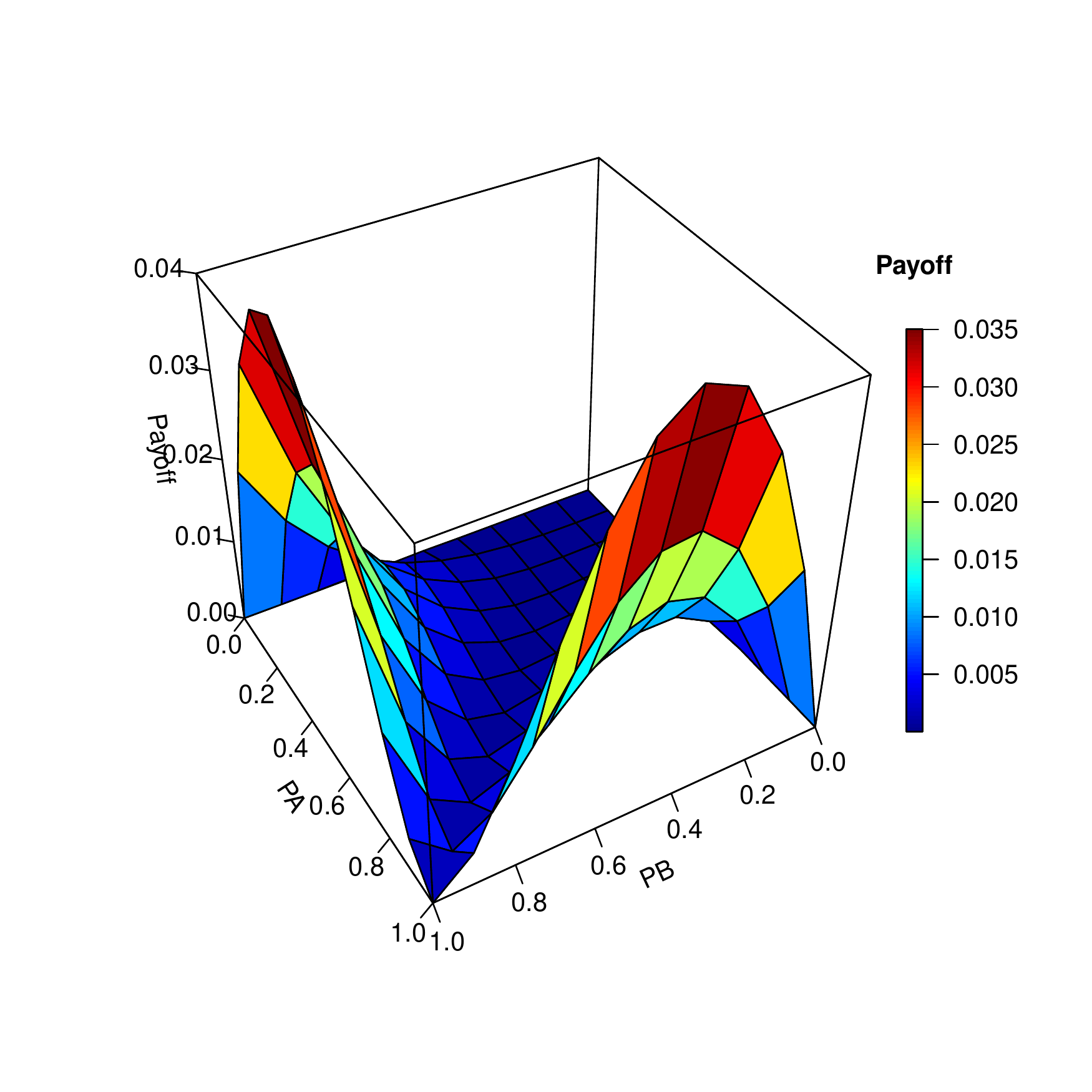}\label{subfig:2-1-1}}\\
\subfloat[$D=AAABB$]{\includegraphics[width=5cm,height=5cm]{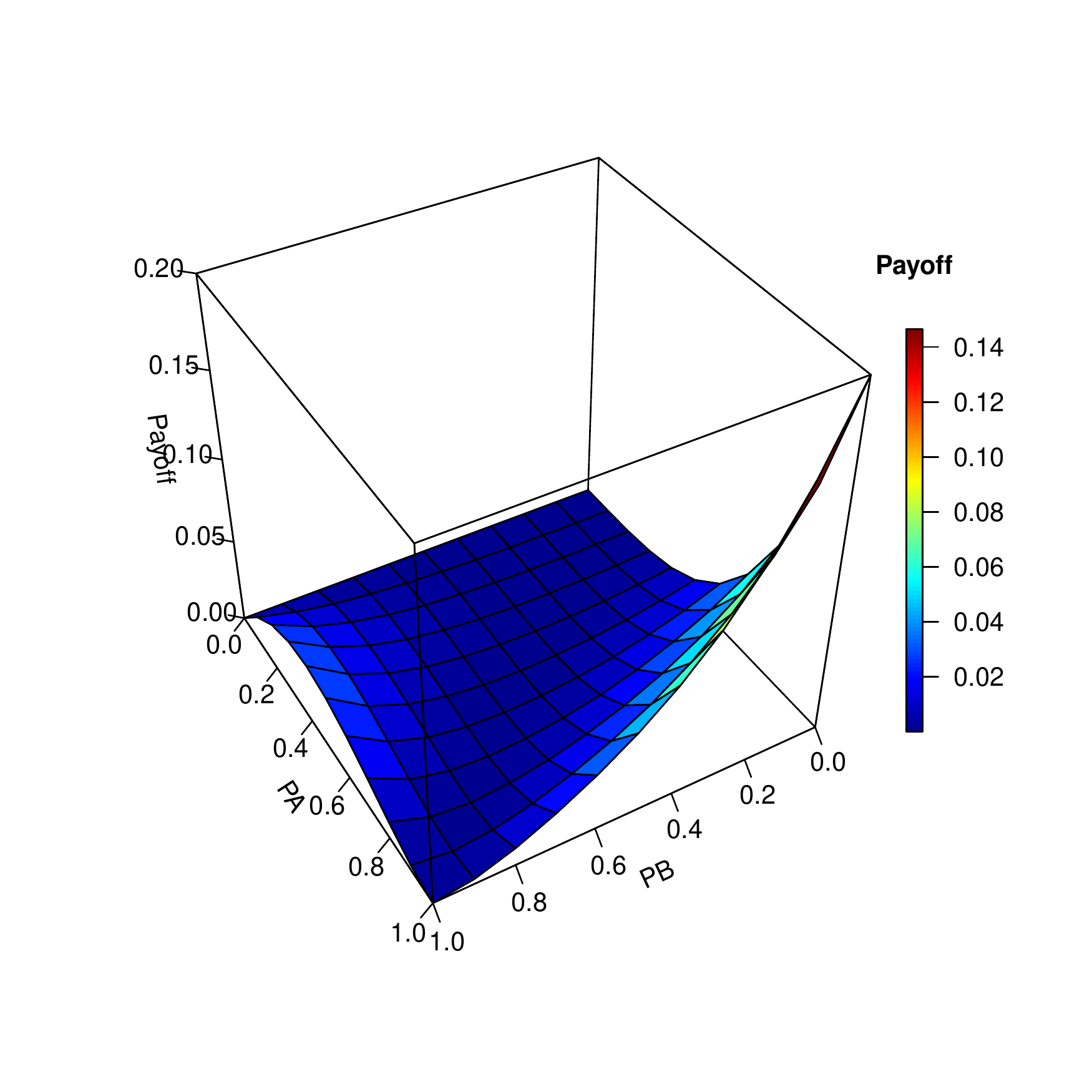}\label{subfig:3-1-1}
}\hspace{20pt} \subfloat[$D=A\cdot\cdot  A B \cdot\cdot B A\cdots \cdots ABBB$]{\includegraphics[width=5cm,height=5cm]{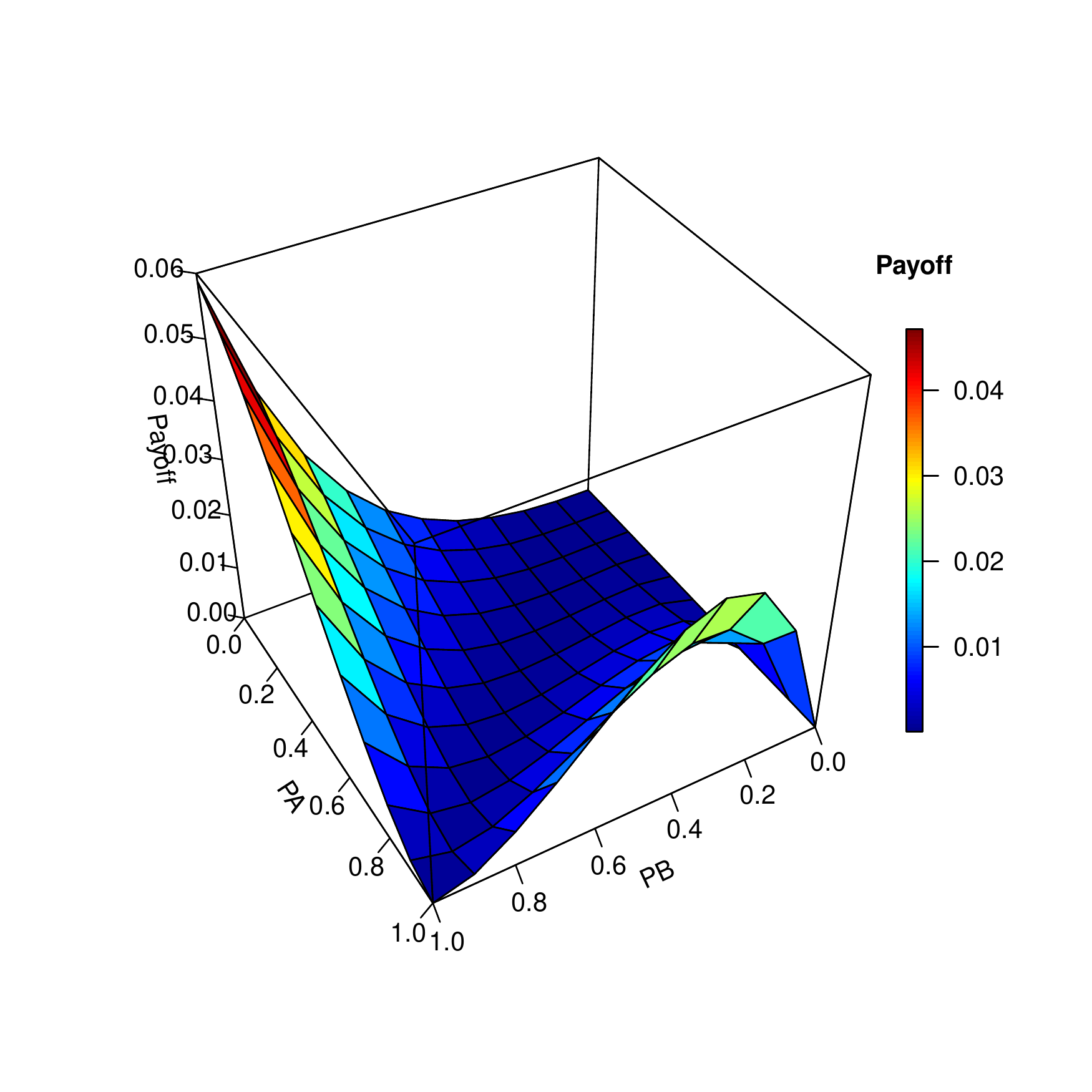}\label{subfig:4-1-1}
}\caption{\label{MC3D}The payoffs across varying winning probability of Arms
$A$ and $B$ under four different nonrandom strategies.}
\end{figure}

 Due to $Q$ in the theoretical is also related to the probability distribution of the two arms, the impact of $Q$ on profits should also be considered when adjusting the distribution of two arms. Fig. \ref{MC3D} shows the three-dimensional image of casino's payoff across the various winning probability of the two arms, which vividly elucidates that the casino could select the  the winning probabilities of $A$ and $B$ to gain the largest profit.
 
\begin{figure}
\centering 
\subfloat[$D=AB$]{\includegraphics[width=5cm,height=4cm]{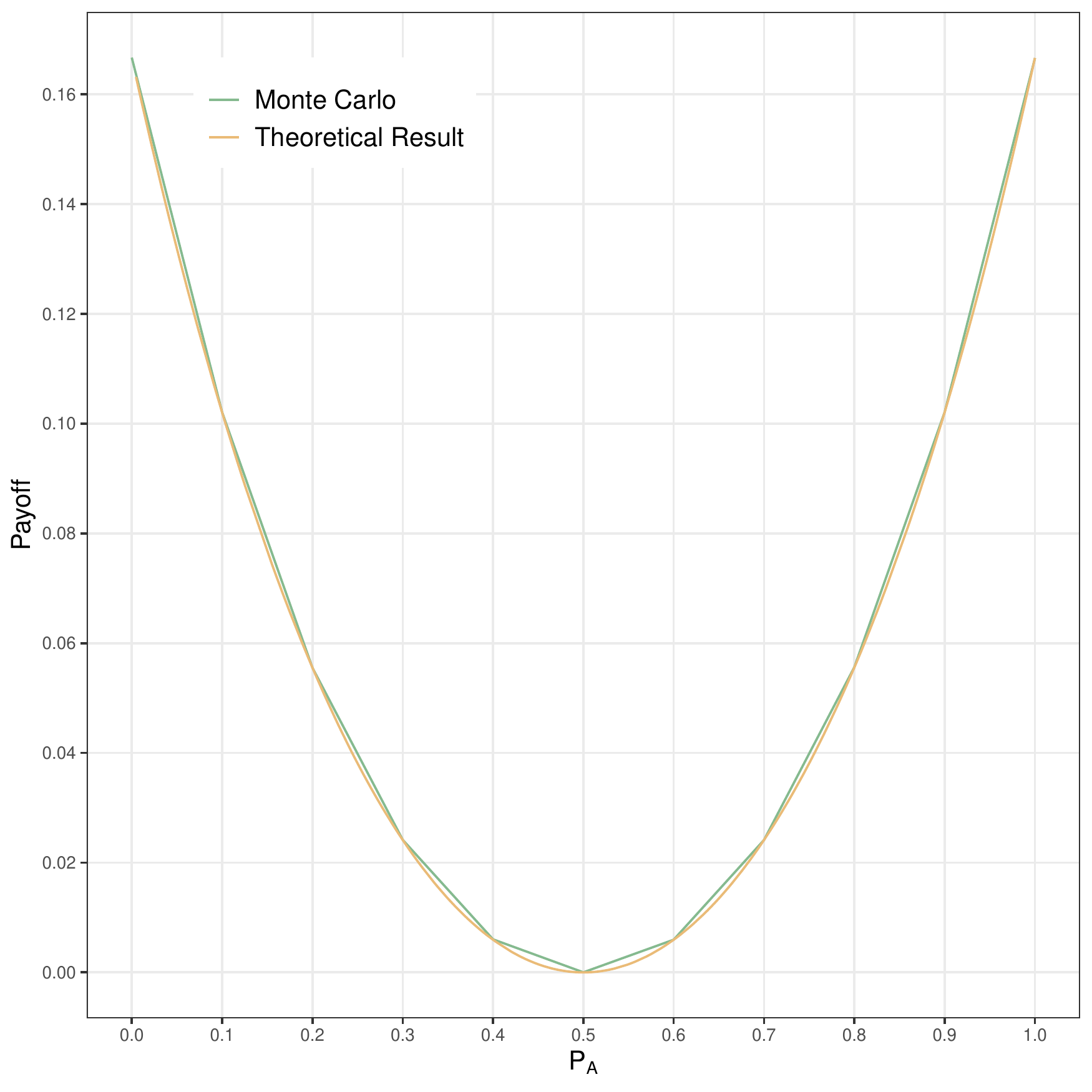}\label{subfig:1}}\hspace{30pt} \subfloat[$D=AABB$]{\includegraphics[width=5cm,height=4cm]{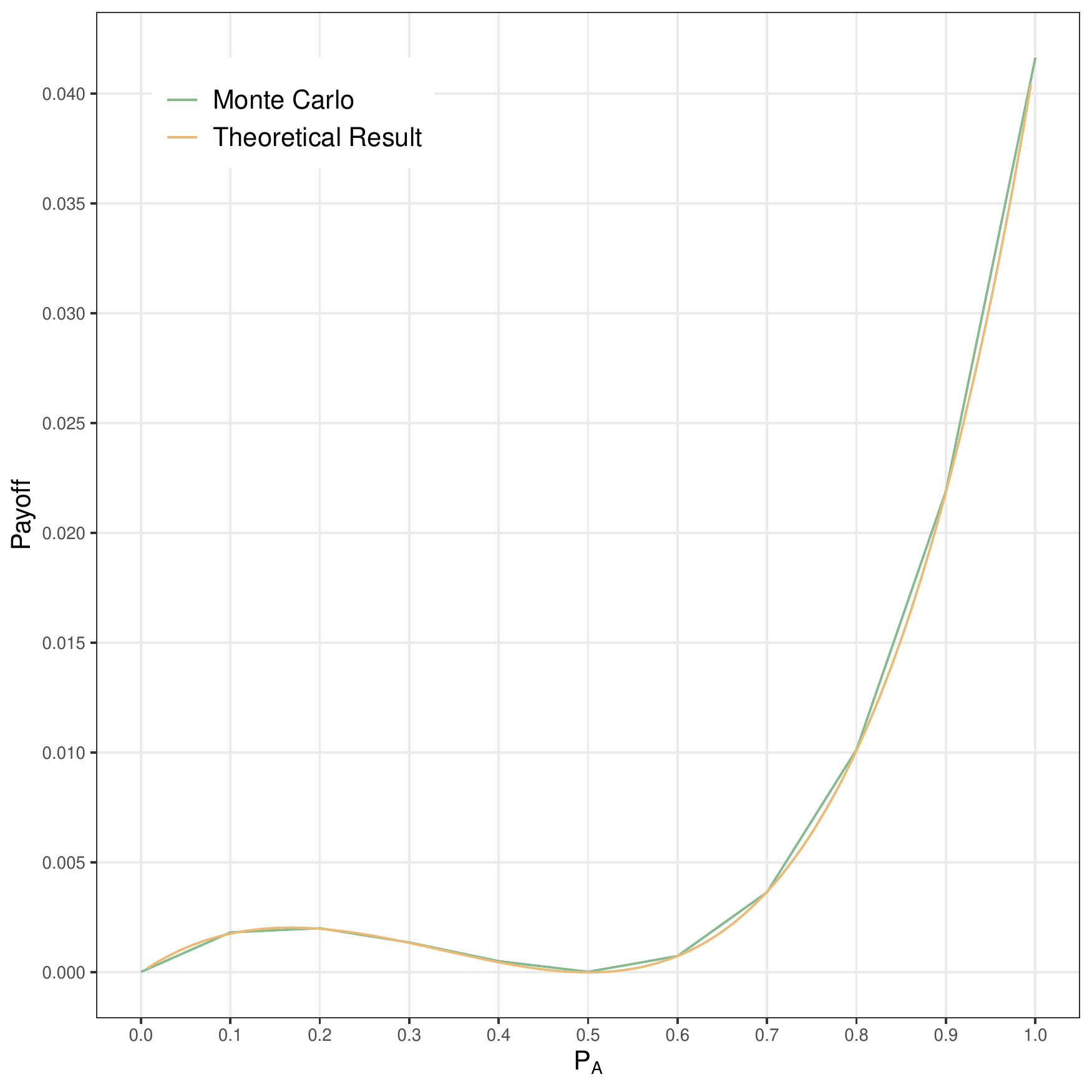}\label{subfig:2}}\\
\subfloat[$D=AAABB$]{\includegraphics[width=5cm,height=4cm]{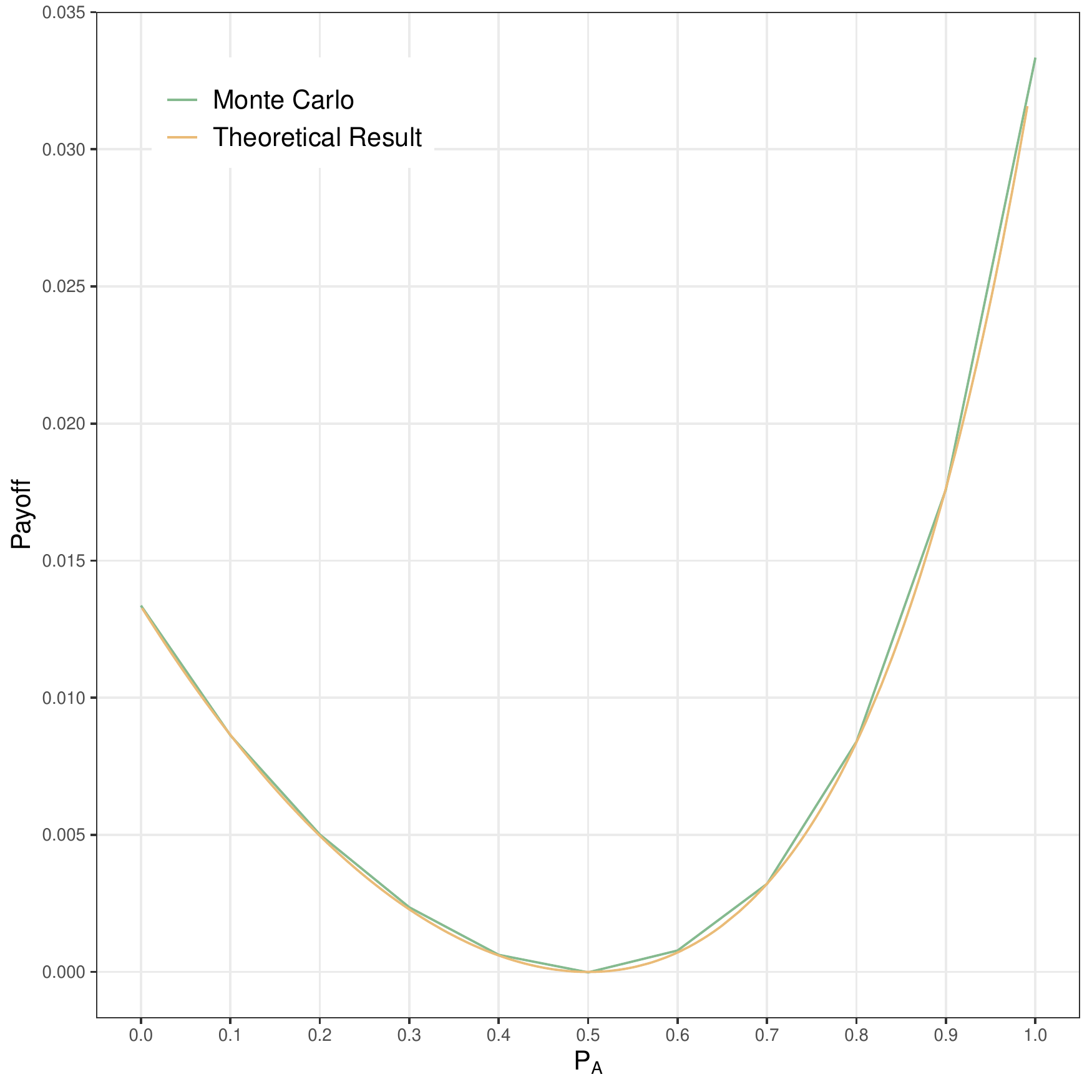}\label{subfig:3}}\hspace{30pt} \subfloat[$D=A\cdot\cdot  A B \cdot\cdot B A\cdots \cdots ABBB$] {\includegraphics[width=5cm,height=4cm]{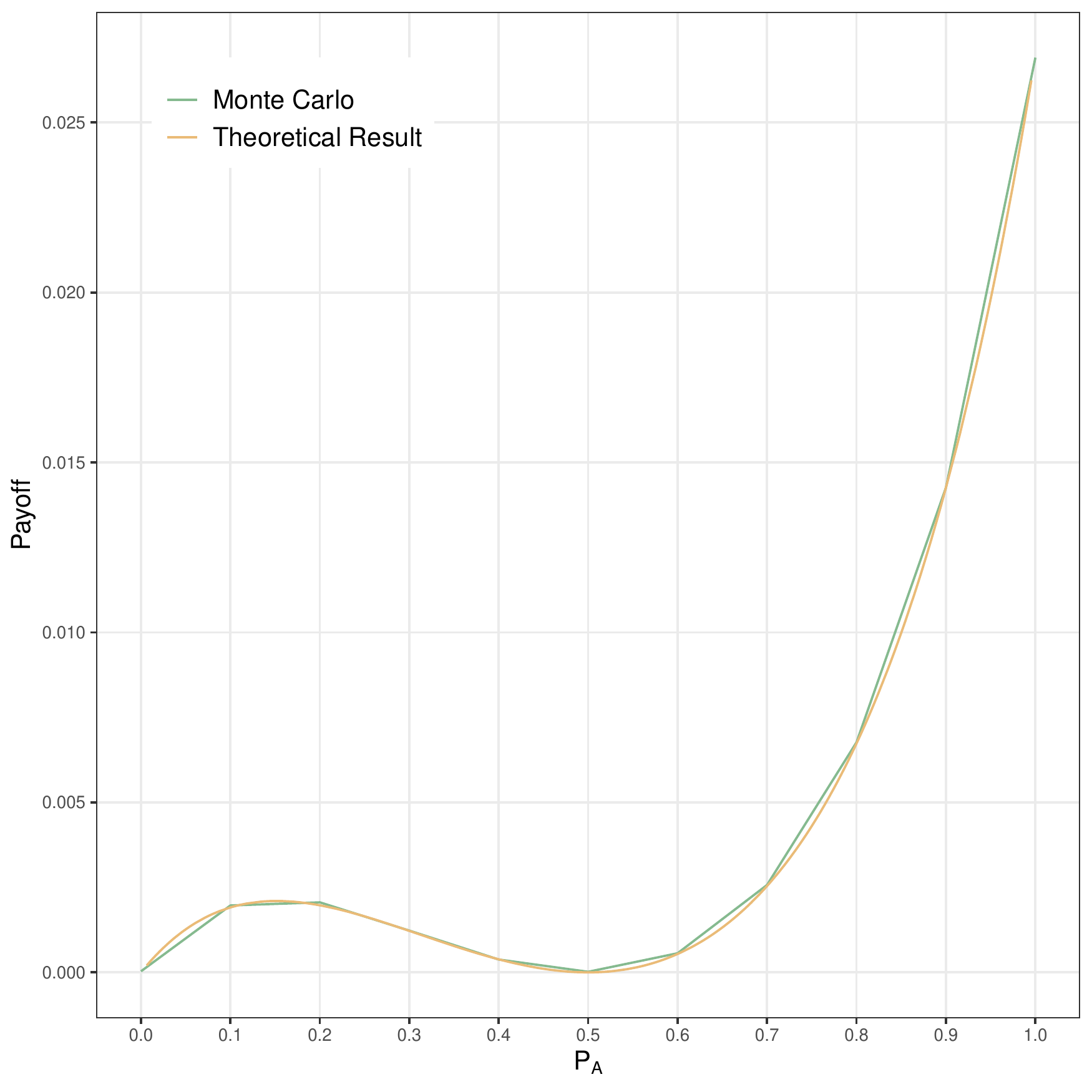}\label{subfig:4}}
\caption{\label{2Dplot}The sample mean payoff of casino across probability of Arm $A$, under the fixed probability of Arm $B$, $p_{B}=0.5.$ }
\end{figure}

 Next, we aim to compare the theoretical payoff and its simulation results by cutting four slice of the three-dimensional surface in Fig. \ref{MC3D}. Without loss of generality, we fixed the probability of Arm B with $p_{B}=0.5$. Fig. \ref{2Dplot} shows the theoretical and simulation's curves for those four nonrandom strategy, which obviously implies that the theoretical conclusions are highly consistent with the simulated results  and then verifies Theorem 1.

\subsection*{A empirical study with a real two-armed futurity slot machine}\label{secs32}

This part considers the antique Mills Futurity slot machine, designed in 1936 by the Chicago Mills Novelty Company\cite{Ethier2010} . There are two screens on the slot machine, one screen recording the current number of consecutive failures and the other screen displays the current mode. In detail, a player consumes 1 coin per coup and 
when the number of consecutive failures  reaches $J=10$, all 10 coins will be returned at this time,  but this experiment still considers $J=2$ extremely attracting the players. Its internal structure is a periodic cam with Mode E or Mode O, corresponding to the Arm $A$ and Arm $B$ and essentially both follows multi-points distribution shown in Table \ref{Tab1}. Both modes have distinct winning probabilities and rewards and the two-armed machines are ``fair'' in the sense that either of the arms has a 50\% chance of making the player profitable, advertised by the casino. However, the casino does not allow you to play one arm and this experiment would check the trick that alternatively using the two ``fair'' arms will make money for the casino.

\begin{figure}
\centering 
\subfloat[$D=AB$]{\includegraphics[width=5cm,height=4cm]{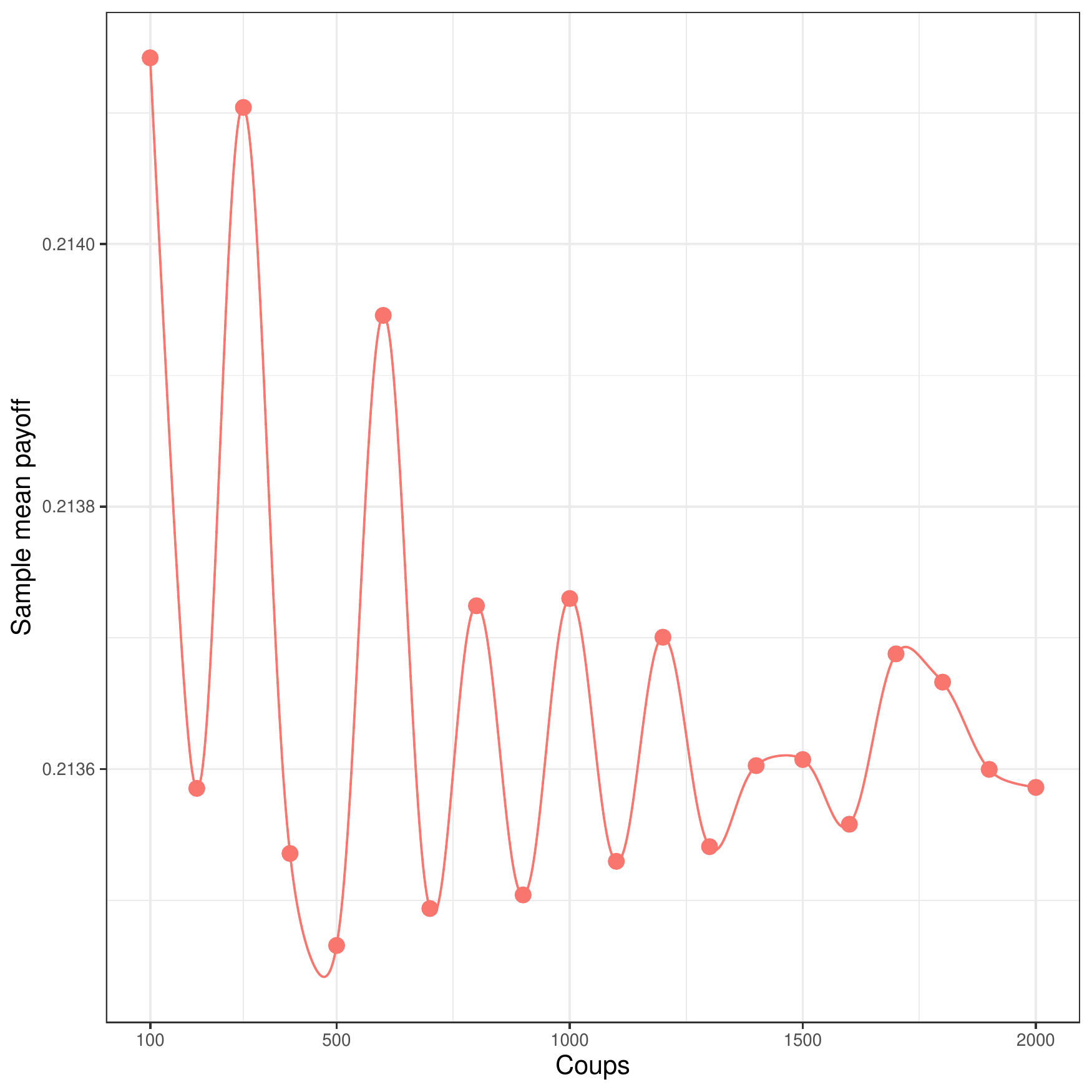}\label{subfig:1}}\hspace{30pt} \subfloat[$D=AABB$]{\includegraphics[width=5cm,height=4cm]{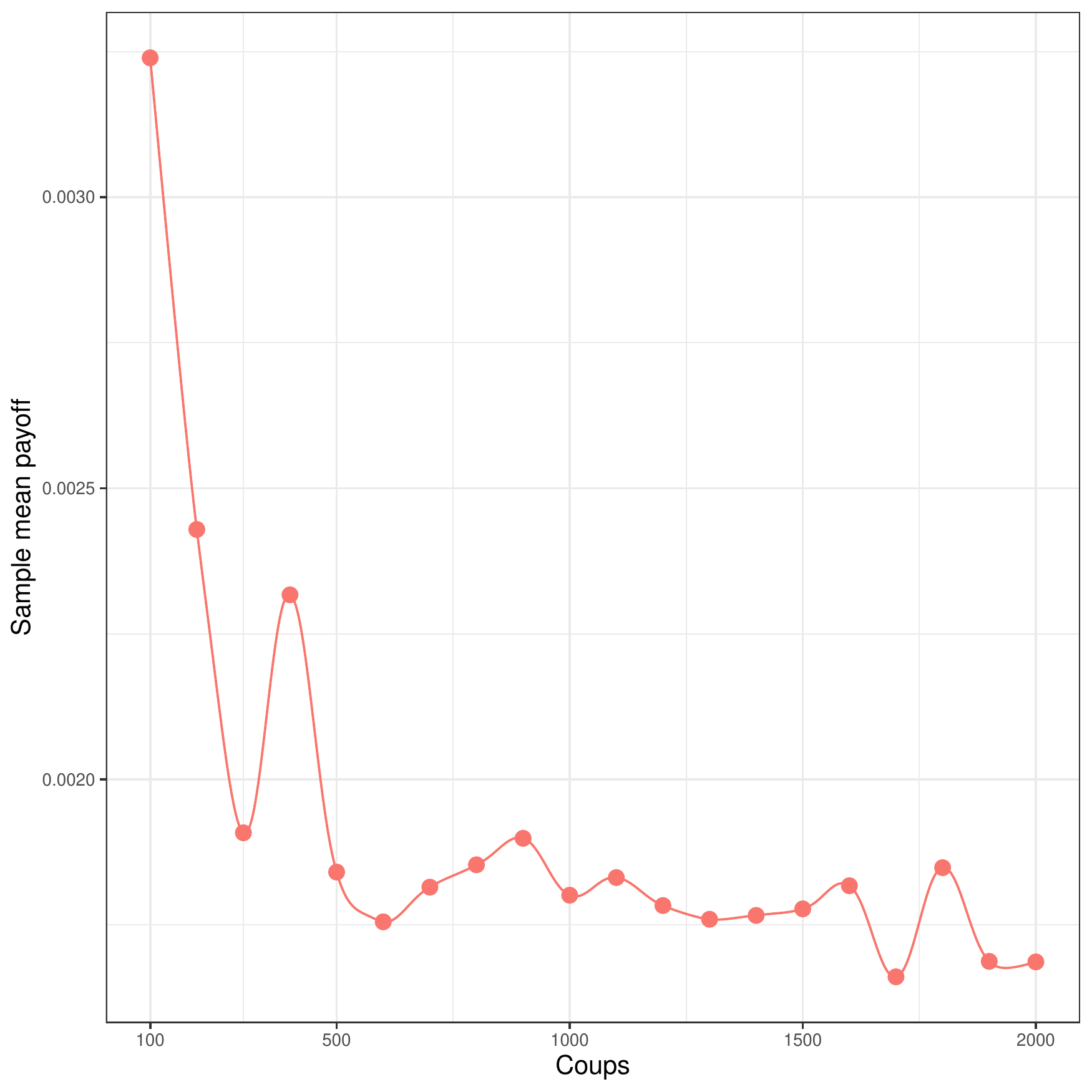}\label{subfig:2}}\\
\subfloat[$D=AAABB$]{\includegraphics[width=5cm,height=4cm]{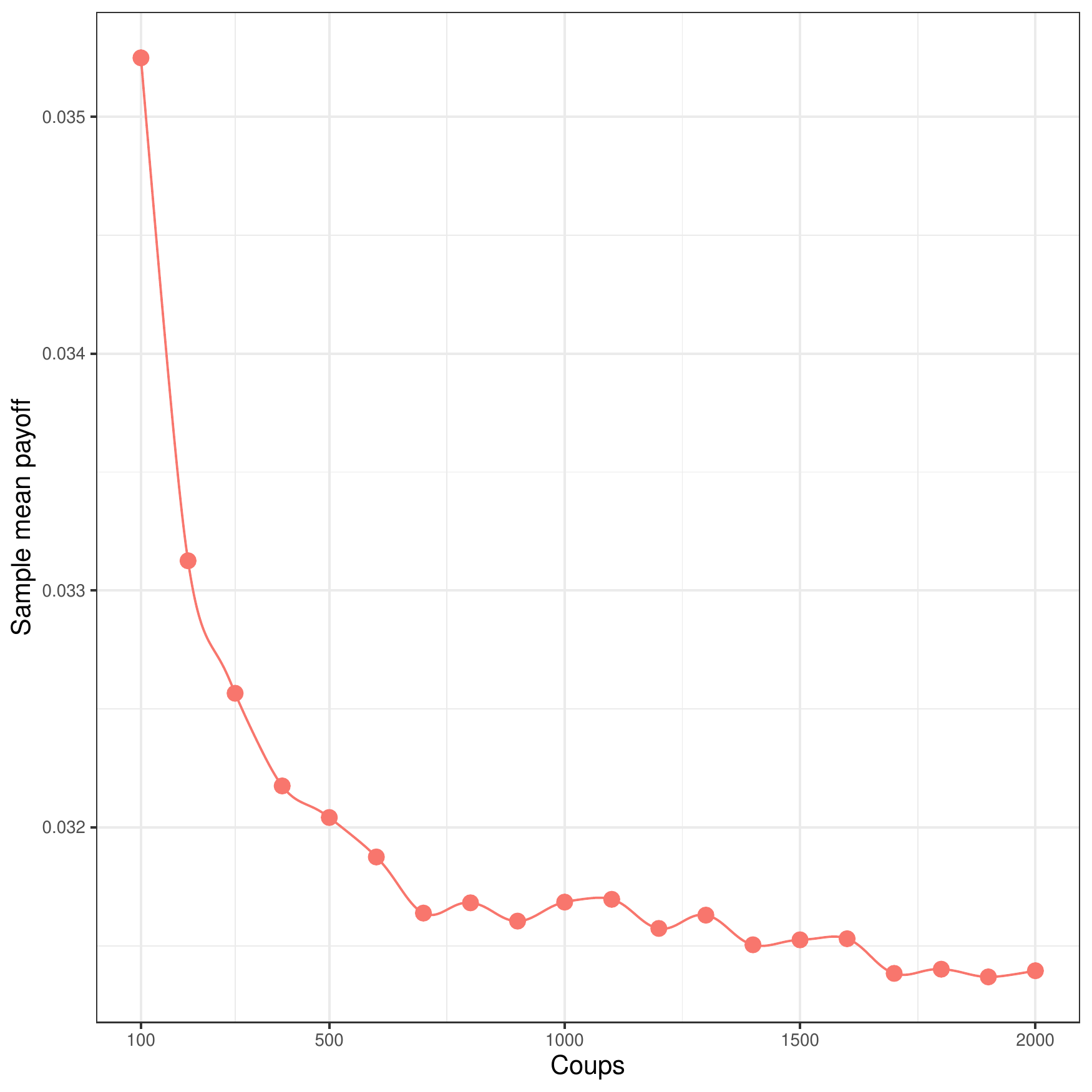}\label{subfig:3}}\hspace{30pt} \subfloat[$D=A\cdot\cdot  A B \cdot\cdot B A\cdots \cdots ABBB$] {\includegraphics[width=5cm,height=4cm]{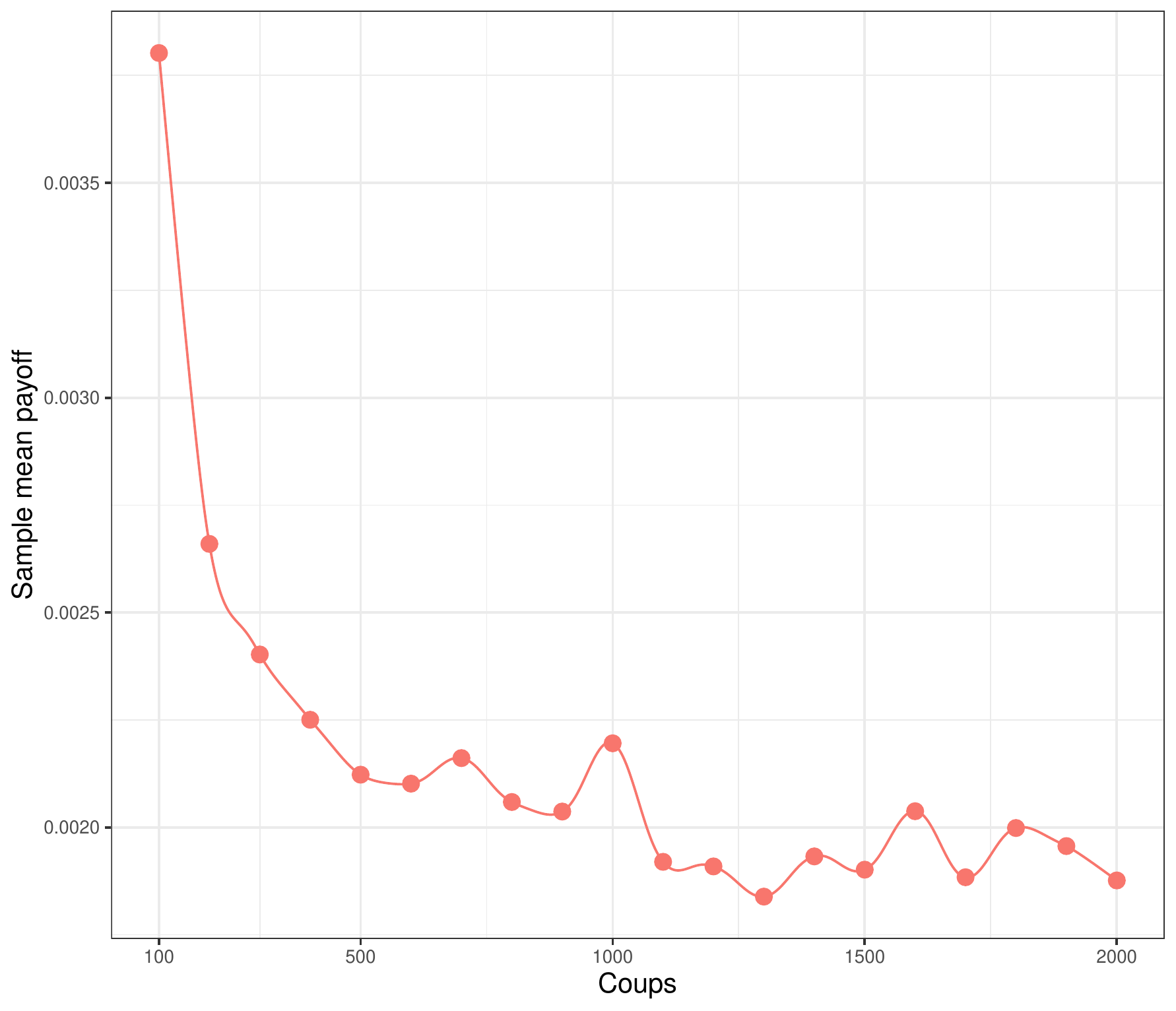}\label{subfig:4}}
\caption{\label{2DFuturity}The cumulative payoff of casino across the number of coups and different strategies $D$ with a two-armed antique Mills futurity slot machine\cite{Ethier2010} .}
\end{figure}

In this application, we transform the multipoint distribution of each mode into two-point distribution. For each mode, we split it into positive payoff and loss and then the reward of each model are obtained, such that the single mode is fair. Particularly, we show the casino's mean profit for the four strategies in Fig. \ref{2DFuturity}, showing that the sample mean profit of casino converges to positive value after long play, which still confirms the conclusion that the casino can always earn
money in the long run under the two-armed Futurity bandit under a equivalent compensation rule that consecutive failures reaching 2.

\begin{table}
\centering
\begin{tabular}{|c|c|c|c|c|c|c|c|}
\hline 
$reward\backslash probability$ & 0 & 3 & 5 & 10 & 14 & 18 & 150\tabularnewline
\hline 
Mode E & 0.968 & 0.003 & 0.007 & 0.018 & 0.004 & 0 & 0\tabularnewline
\hline 
Mode O & 0.357 & 0.576 & 0.064 & 0 & 0 & 0.002 & 0.001\tabularnewline
\hline 
\end{tabular}
\caption{\label{Tab1}The multipoint distribution of reward in Mode E and Mode O of the futurity slot machine.}
\end{table}

\section*{Discussion}\label{sec4}
This article implies that the root cause of losing games for gamblers rests with 
the sufficiently mathematical logic  of any gambling equipment and the extremely sophisticated program design based on probability modeling and random calculation.
This work rigorously demystifies the so-called casino loyalty programmes designing fair returns with one-armed Futurity bandit to attract the gamblers and the hoax making a buck from the gamblers continuously by using two-armed Futurity bandit, which implies the seemingly fair fraud of two-armed Futurity bandit. Explicit mathematical expression
of casino profits has been concluded in the results and the corresponding images vividly elucidate the functional change along with the considered parameters, which also implies that the the casino can be always profitable in the long run. Experiments are designed to check the correctness of the theoretical results by simulation and a real  two-armed futurity slot machine with more complex output has also been checked to verify the conclusion. We anticipate the assay to be a benefit for the players to thoroughly recognize the gambling industry who usually design the so-called loyalty programme with fairness to attain the profitable purpose. However, we have no wish for using the theoretical findings to design the slot machine by the casinos, even by other businesses such as
discount marketing, bundled sales or other induced consumption tactics.
This article may be a starting point for studying the profit achievement of casinos with
more sophisticated multi-armed Futurity bandits based on the probabilitic tool.  

\section*{Online content}
Any methods, additional references, nature research reporting summaries, supplementary information, acknowledgements, peer review information; details of
author contributions and competing interests; and statements of
data and code availability are available at online.

\newpage


\newpage

\section*{Method}
 Below we will uncover the secrets behind the long-term profitability of this slot machine step by step. The player will always be interested in the question of why casinos can claim that every arm is fair, which is what attracts the player to this game. Although the strategy $D$ provided by the player has many results, it seems that the player can formulate a strategy that is beneficial to him in advance, but in fact the casino will treat many of the player's strategies as the same strategy. So what is the relationship between the different strategies? After the casino gets the strategy $D$, it needs to find out how different strategies affect the casino's profit. This slot machine can also be regarded as a confrontation game between the casino and the player, so how is the asymptotic profit expectation difference between the casino and the player calculated? After the casino calculates the asymptotic profit expectation, we can still ingeniously calculate whether the value of this profit is strictly positive, which explains why the casino will be profitable in the long run.
\subsection*{Why casinos can claim every arm of the slot machine is fair?}The source of the casino's profit is only 1 coin paid by the player before each game. The player's profit from the slot machine is divided into two parts, one part is the payoff $u$ obtained by winning a single game, and the other part is the refund of the casino obtained by losing two consecutive games. We can choose one arm for analysis. If the player only play the $A$ arm, $p^\circ_A$ represents the asymptotic probability of the player getting the futurity reward per game, then the player's expectation of the asymptotic revenue per game is $\mu^ *_A=p_Au_A+2p^\circ_A$ where $u_A$ is the payoff the player gets from winning a single game. The casino can make $\mu^*_A=1$ by adjusting the parameters $p_A$ and $u_A$. This means that the player's asymptotic payoff expectation per game is equal to the 1 coin payoff received by the casino before each game. S.N Either and Jiyeon Lee calculated the value of $p^\circ_A$, that is, $p^\circ_A=\frac{p_Aq^2_A}{1-q^2_A}$ where $q_A=1-p_A$. Then in order to ensure fairness, the casino needs to ensure $u_A=\frac{3-2p_A}{2-p_A}$ while modifying the arm's payoff distribution. In the same way, the casino can also make the $B$ arm fair, but the set parameters need to be $p_A\neq p_B$, $0<p_A,p_B<1$.
\subsection*{What is the relationship between different non-random mixing strategies?}Fixing the general $r, s>0$, no matter what strategy $D$ the player offers, the casino only cares how this strategy affects the casino's asymptotic profit expectations. We denote $p^\circ_D$ as the asymptotic probability of player getting the futurity reward under strategy $D$ per game, and $p^D_i$ denotes the winning probability of the $i$th game under strategy $D$. S.N Either and Jiyeon Lee preliminarily gave the form of $p^\circ_D$. On this basis, we preliminarily calculated the casino's asymptotic profit expectation and the value of $p^\circ_D$.
\begin{lemma}
The casino's asymptotic profit expectation\\
$$R=2(p^\circ_D-\frac{r}{r+s}p^\circ_A-\frac{s}{r+s}p^\circ_B)$$
where $$p^\circ_D=\frac{1}{r+s}\sum\limits_{k=1}^{r+s}\big(\sum\limits_{j=1}^{r+s}p^D_j\prod\limits_{i=j+1}^{j+2k}q^D_i\big)\frac{1}{1-(q^r_Aq^s_B)^2}$$
where $q^D_i=1-p^D_i$.
\end{lemma}
From the above lemma, we can observe that different strategies only affect the value of $p^\circ_D$, and the value of $p^\circ_D$ will affect the casino's asymptotic profit expectation $R$. From the expression of $p^\circ_D$, we observe that the value of $p^\circ_D$ is equal between some strategies.
\begin{lemma}
Fix the general r, s and l, where $l=1,2,...,r+s$. Consider nonrandom-pattern strategies $D_1$ and $D_2$, where $p^{D_1}_i=p^{D_2}_{i+l}$ for any $i=1,2,...,r+s$. Then $p^\circ_{D_1}=p^\circ_{D_2}$
\end{lemma}
To understand the above lemma more intuitively, we can think of steps $A$ and $B$ in the strategy as $r$ $A$ balls and $s$ $B$ balls. These balls are placed one by one on a round table, then the value of $p^\circ_D$ for the same arrangement is equal. For example, if $r=4$, $s=2$, the following two arrangements are the same value of $p^\circ$, that is $p^\circ_{AABABA}=p^\circ_{ABABAA}$\\

\qquad
$
\begin{matrix}
  \  & \  & A & \longrightarrow & A  & \  & \  \\
  \  & \nearrow & \  & \  & \  & \searrow & \  \\
  A & \  & \  & \  & \  & \  & B \\
  \  & \nwarrow & \  & \  & \  & \swarrow & \  \\
  \  & \  & B & \longleftarrow & A & \  & \
\end{matrix}
$
\qquad \qquad
\qquad
\qquad
$
\begin{matrix}
  \  & \  & A & \longrightarrow & B  & \  & \  \\
  \  & \nearrow & \  & \  & \  & \searrow & \  \\
  A & \  & \  & \  & \  & \  & A \\
  \  & \nwarrow & \  & \  & \  & \swarrow & \  \\
  \  & \  & A & \longleftarrow & B & \  & \
\end{matrix}
$\\
In this way, any strategy provided by the player can be regarded by the casino as a strategy starting from Arm $A$ in the process of calculating the profit. And use the vector $\boldsymbol{a}(h,r,s)$ to represent the structure of this strategy, that is (\ref{eq1}). Where $h$ is the number of times the arm is switched in a single strategy. We conjecture that in the strategy for fixed $r$,$s$ provided by the player, the more frequently switching arms, the higher the profit of the casino, that is, $R$ and $h$ are positively correlated.
\subsection*{How is the casino's asymptotic profit expectation calculated?} We notice that only the value of $p^\circ_D$ is the factor that can affect the profit of the casino. From the above form of $p^\circ_D$, we need to analyze the arrangement of each step in strategy $D$ and the next $2k$($k=1,2,...,r+s$) steps. This calculation seems overly complicated as the strategy varies, even though we have simplified the analysis of strategy $D$. But for the simplest policy $D_1=AA...ABB...B$ i.e. $\boldsymbol{a}=(r,s)$, we are not that complicated to calculate because it has the least number of different cases between adjacent steps. Let $q_A=x$, $q_B=y$, then we have
$$R_{D_1}=2S(1-(-1)^rx^r)(1-(-1)^sy^s)$$
\begin{lemma}\label{le3}
For any $h\in\mathbb{Z}^*$, consider two nonrandom-pattern strategies with pattern $D$, $D'$.
  $$D=\underbrace{A...A}_{r_1}\underbrace{B...B}_{s_1}...\underbrace{A...A}_{r_{h-1}}\underbrace{B...B}_{s_{h-1}}\underbrace{A...A}_{r_h}\underbrace{B...B}_{s_h}$$  $$D'=\underbrace{A...A}_{r_1}\underbrace{B...B}_{s_1}...\underbrace{A...A}_{r_{h-1}}\underbrace{B...B}_{s_{h-1}}\underbrace{B...B}_{s_h}\underbrace{A...A}_{r_h}$$
 where $1\le k\le h$, $r_k>0$ , $s_k>0$, $\sum\limits_{k=1}^{h}r_k=r$, $\sum\limits_{k=1}^{h}s_k=s$
Then
$$R_D-R_{D'}=2S(1-{b_{2h-1}})(1-{b_{2h}})\big(\sum\limits_{j=0}^{2h-3}(-1)^{j}\prod\limits_{i=1}^{j}{b_{i}}+\sum\limits_{j=1}^{2h-2}(-1)^{j}\prod\limits_{i=1}^{j}{b_{2h-1-i}}\big)$$
where $\boldsymbol{a}=(r_1,s_1,r_2,s_2,...,r_h,s_h)$
\end{lemma}

We observe that strategy $D'$ is obtained by exchanging the strategy segment in $D$ with only $A$ with the immediately following strategy segment with only $B$. The reason for choosing these two special strategies is that most of the product terms $q^D_i$ in the expression of $p^\circ_D$ are equal to $q^{D'}_i$ in the expression of $p^\circ_{D'}$. Due to the symmetry of $A$ and $B$, we can also exchange the strategy segment with only $B$ in $D$ with the immediately following strategy segment with only $A$, and we can get similar results. As can be seen from the previous section, any strategy provided by the player can always be written in the form (\ref{eq1}).
We first exchange the strategy segment $B$ of the  $2h-2$-th segment and the strategy segment $A$ of the $2h-1$-th segment to obtain a new strategy $D_{h-1}$. Then exchange the strategy segment $B$ of the $2h-4$-th segment and the strategy segment $A$ of the $2h-3$-th segment in $D_{h-1}$, so that we can get a new strategy $D_k$, that is
$$D_{k}=\underbrace{A...A}_{r_1}\underbrace{B...B}_{s_1}...\underbrace{A...A}_{r_{k-1}}\underbrace{B...B}_{s_{k-1}}\underbrace{A...A}_{r_k+r_{k+1}+...+r_{h}}\underbrace{B...B}_{s_k+s_{k+1}+...+s_{h}}$$
We keep exchanging until we finally get $D_1=AA...ABB...B$, which happens to be the strategy for which we calculated the casino's asymptotic profit expectations in advance. The difference between the profits of each exchange of the previous strategy can be calculated by the lemma \ref{le3}, so that we sum them all up and prove by induction, we get the casino asymptotic profit expectation $R=2QS$.

\subsection*{Why casinos can be profitable in the long run?}In theorem \ref{th1}, we note that if $x\neq y$, $0<x,y<1$, then $S>0$. Then we only need to observe the positive and negative of $Q$. We can think of $b_1$, $b_2$,...,$b_{2h}$ as $2h$ points and draw them on the round table in clockwise order and we number the points in the order of $\{b_k\}$'s numbering. For the convenience of the explanation of the following formula, we find a representative example $\boldsymbol{a}=(2,1,2,2,1,2,1,2,2,1)$. We first pick out $k$ that satisfy $b_k<0$ and renumber them, they represent the odd number of $A$ or $B$ in the original strategy segment, we call these points ``negative'' points, the rest are called ``positive'' points. Define a function $c_i(j)$, $i,j\in \mathbb{Z}$ that satisfies the following properties:\\
\begin{itemize}
\item $c_1:=\min\{i\mid i>0,b_{i}<0\}$\\
\item $c_k:=\min\{i\mid i>c_{k-1},b_{i}<0\}$\\
\item $c_{\delta}:=\max\{c_k\mid c_k\le 2h\}$\\
\end{itemize}

Now we consider the case where $\delta>1$. In our example we try to relabel elements where $b_k<0$ as follows\\

$
\begin{matrix}
  \  & \  & b_2 & \longrightarrow & b_3 & \longrightarrow & b_4 & \  & \  \\
  \  & \nearrow & \  & \  & \  & \ & \ & \searrow & \  \\
  b_1 & \  & \  & \  & \  & \  & \ & \ & b_5 \\
  \uparrow & \  & \  & \  & \  & \  & \ & \ & \downarrow\\
   b_{10} & \  & \  & \  & \  & \  & \ & \ & b_6 \\
  \  & \nwarrow & \  & \  & \  & \ & \ & \swarrow & \  \\
  \  & \  & b_9 & \longleftarrow & b_8 &\longleftarrow & b_7 & \  & \
\end{matrix}
$
\qquad 
$
\begin{matrix}
  \  & \  & \boldsymbol{b_{c_1}} & \longrightarrow & b_3 & \longrightarrow & b_4 & \  & \  \\
  \  & \nearrow & \  & \  & \  & \ & \ & \searrow & \  \\
  b_1 & \  & \  & \  & \  & \  & \ & \ & \boldsymbol{b_{c_2}} \\
  \uparrow & \  & \  & \  & \  & \  & \ & \ & \downarrow\\
   \boldsymbol{b_{c_4}} & \  & \  & \  & \  & \  & \ & \ & b_6 \\
  \  & \nwarrow & \  & \  & \  & \ & \ & \swarrow & \  \\
  \  & \  & b_9 & \longleftarrow & b_8 &\longleftarrow & \boldsymbol{b_{c_3}} & \  & \
\end{matrix}
$\\

We observe that $Q$ is formed by the sum of products of different numbers of $b_k$ that are adjacent. We factor the product term in the following two ways. First we can choose $i$ as the ``starting'' point and $i+j$ as the ``end'' point, and a formula is generated from this, that is, $\xi(i,i+j):=(-1)^{j+1}\prod\limits_{k=i}^{i+j}{b_k}$ for $0\le j< 2h$ and $\xi(i,i-1):=1$. And if the end point is positive point, that is $b_{i+j}>0$, then the first factorization method is $\xi(i,i+j)=-\xi(i,i+j-1)b_{i+j}$. If we find two negative points $c_i$ and $c_{i+j}$ as starting and end points, and similarly a formula is generated from this, that is, $\eta(i,i+j)=\xi(c_i,c_{i+j})$ for $0\le j <\delta$. And we can obtain the following lemma by proof by contradiction.
\begin{lemma}\label{le4}
For $\delta>1$, $0<j<\delta$, if $\eta(i,i+j)<0$, then we have $m(i,i+j)$ or $n(i,i+j)$ satisfies $\eta(i,i+m)<0$ or $\eta(i+n,i+j)<0$ for some $0<m \le j$ or $0\le n < j$ and $m$, $j-n$ are odd. Meanwhile it satisfies the following properties:
\begin{itemize}
\item There is no $0< m_0 <m$, $0< n_0 <m$ so that $\eta(i,i+m_0)<0$ or $\eta(i+n_0,i+m)<0$.\\
\item There is no $n< m'_0< j$, $n< n'_0< j$ so that $\eta(i+n,i+m'_0)<0$ or $\eta(i+n'_0,i+j)<0$.
\end{itemize}
\end{lemma}
Then the second factorization method is $\eta(i,i+j)=-\xi(c_i,c_{i+n}-1)\lvert\eta(i+n,i+j)\rvert$, where $n$ is found by the above lemma and it is assumed that such $n$ exists. For example $\eta(1,3)=-\xi(2,4)\lvert\eta(2,3)\rvert<0$, where $\xi(2,4)>0$ and $\eta(2,3)<0$. This way if we fix the starting point $i$, some product terms starting from $i$ can be written as a polynomial. According to the above two factorization methods, we start the factorization from the $\xi(i,i+2h-1)$ with the most product terms, that is, the end point $i+2h-1$(i.e $i-1$) farthest from the starting point. And record the new end point obtained by factoring, and then continue to factorize according to the above methods from the new end point until it cannot be decomposed. The end point at this 
time must be a negative point. Then we select the farthest end point that did not appear just now, and continue to decompose according to the above method, so that the sum of all product terms starting from $b_i$ can be written as the sum of several polynomials. Going back to our example, the sum of the product terms starting at $b_2$ can be reduced to the sum of the following polynomials.
\begin{align*}
&\sum\limits_{j=0}^{9}\xi(2,2+j)\\
=&-b_2\big(1-b_3(1-b_4(1-b_5b_6b_7(1-b_8(1-b_9)))\big)+\xi(2,5)(1-b_6)+\xi(2,10)(1-b_1)
\end{align*}

We notice that the value of the second term of the product in the above summation formula is between 0 and 1, that is, when we study the positive and negative of the product term, we only need to study the positive and negative of the first term of the product. For this part, we noticed that its end point can no longer be factored according to the above methods, but we can try to factorize its starting point, and we also have the following two methods.
\begin{itemize}
\item If $i$ and $i+j$ as starting and end points, and if $b_i>0$, then the third factorization method is $\xi(i,i+j)=-b_i\xi(i+1,i+j)$.\\
\item If $c_i$ and $c_{i+j}$ as starting and end points, the fourth factorization method is $\eta(i,i+j)=-\lvert\eta(i,i+m)\rvert\xi(c_{i+m}+1,c_{i+j})$ where $m$ is found by the above lemma and it is assumed that such $m$ exists.
\end{itemize}

For the function $Q$, we have
$$Q=\sum\limits_{l=1}^{h}(1-{b_{2l}})\big(\sum\limits_{j=0}^{2h-1}\xi(2l+1,2l+j)\big)$$
We first take all negative points as endpoints, and observe the addition between two adjacent endpoints. Without losing generality, we observe the addition between $c_i$ and $c_{i+1}$. According to the new form of $Q$, we consider partial summation
$\sum\limits_{c_i<2l<c_{i+1}}^{}(1-{b_{2l}})\big(\sum\limits_{j=0}^{2h-1}\xi(2l+1,2l+j)\big)$.
In addition, on the basis of the original sum, if $c_{i+1}$ is even, we also need to consider $-{b_{c_{i+1}}}\big(\sum\limits_{j=0}^{2h-1}\xi(c_{i+1}+1,c_{i+1}+j)\big)$. Also, if $c_{i}$ is even, we also need to consider $\sum\limits_{j=0}^{2h-1}\xi(c_{i}+1,c_{i}+j)$. Then according to the different parity of $c_i$, $c_{i+1}$, it can be divided into four addition cases. It just so happens that the four segments summation formula that our example is divided into corresponds to four different cases.
$$
\begin{matrix}
case& 1:&\boldsymbol{b_{c_1}} & \longrightarrow & b_3 & \longrightarrow & \boldsymbol{b_4} & \longrightarrow & b_{c_2}   \\
\\
case &2:&b_{c_2} & \longrightarrow & \boldsymbol{b_6} & \longrightarrow & b_{c_3} & \   \\
\\
case &3:&b_{c_3} & \longrightarrow & \boldsymbol{b_8} & \longrightarrow & b_9 & \longrightarrow & \boldsymbol{b_{c_4}}   \\
\\
case &4:&\boldsymbol{b_{c_4}}& \longrightarrow & b_1 & \longrightarrow & \boldsymbol{b_{c_1}} & \   \\
\end{matrix}
$$
The end point $c_4$ of the line in case 3 is exactly the starting point of the line in case 4, and the end-to-end connection is unique, then according to the decomposing method we discussed earlier, $Q$ can be divided into the above four cases. After the above discussion, all negative polynomials starting from $i$ and $b_i>0$ can always find positive polynomials starting from $i+1$. According to the third factorization method, the sum of the two is positive. Let's take case 1 as an example.  
\begin{align*}
&\sum\limits_{j=0}^{9}\xi(3,2+j)+(1-b_4)\sum\limits_{j=0}^{9}\xi(5,4+j)\\
=&1-b_3\big(1-b_4(1-b_5b_6b_7(1-b_8(1-b_9)))\big)+\xi(3,5)(1-b_6)+\xi(4,10)(1-b_1)(1-b_3)\\
+&\big(1-b_5b_6b_7(1-b_8(1-b_9(1-b_{10}b_1b_2)))\big)(1-b_4)-b_5(1-b_6)(1-b_4)+\xi(5.10)(1-b_1)\\
>&1-b_3(1-b_4(1-b_5b_6b_7(1-b_8(1-b_9)))+\xi(3,5)(1-b_6)+\xi(5,10)(1-b_1)
    \end{align*}
If there are more than one $l$ in the summation segment satisfying $c_i<2l<c_{i+1}$, then the polynomial sum generated by the starting and end points of the same summation segment is still positive. Now our summation formula is greater than the sum of the negative polynomial starting from $c_{i+1}$ and the positive polynomial starting from $c_i+1$. We also get similar results in the other 3 cases, only the difference of whether or not $\xi(1,2h)$ is included. This way we don't have to consider negative polynomials starting from any positive points.
As we discussed earlier, the first product term that makes up a negative polynomial starts and ends at negative points. Without loss of generality we write this product term as $\eta(i+1,i+j)$. Then according to our factoring method and Lemma \ref{le4}, we must be able to use the fourth factoring method to get $\eta(i+1,i+j)=-\lvert\eta(i+1,i+m)\rvert\xi(c_{i+m}+1,c_{i+j})$. Corresponding to our four summation cases, the negative polynomial generated at the end of the summation segment must have a unique correspondence with the positive polynomial generated at the beginning of some segment, and the addition of the two is positive. In this way, either a closed loop is found so that the sum of all polynomials in the closed loop is positive, or there are no more negative polynomials in a certain summation segment. Then we get $Q>0$. In our example, the negative polynomial $\xi(5,10)(1-b_1)$ of the summation section of case 1 can be found in the summation section of case 3 with its unique corresponding positive polynomial $\xi(8,10)(1-b_1)$, such that the sum of the two is positive, that is $\xi(8,10)(1-b_1)(1-b_5b_6b_7)>0$. Next, the negative polynomial in the summation segment of case 3 can also find the corresponding positive polynomial in the summation segment of case 1, thus forming a closed loop. Also in our example the summation segments of case 2 and case 4 do not contain negative polynomials. Hence we get $Q>0$.

For the relatively simple case $\delta=0$, we can directly concatenate all the expressions into a polynomial, that is
$$Q=\sum\limits_{l=1}^{h}(1-{b_{2l}})\big(1-b_{2l+1}(1-b_{2l+2}(...(1-b_{2h+2l-2}(1-b_{2h+2l-1}))...))\big)>0$$
And for $\delta=1$, without losing generality we suppose that ${b_{m^*}}<0$ and $m^*$ is odd, then
\begin{align*}
Q=&\sum\limits_{l=1}^{h}(1-{b_{2l}})\big((1-b_{2l+1}(...(1-b_{m^*-2}(1-b_{m^*-1}))...))\\
+&\lvert\xi(2l+1,m^*)\rvert(1-b_{m^*+1}(...(1-b_{2h+2l-2}(1-b_{2h+2l-1}))...))\big)>0
    \end{align*}

From this, we have revealed the secret of the long-term profitability of the Futurity slot machine in a mathematical sense.

\section*{Data Availability}
 There are no data underlying this work.
\section*{Code availability}
The DAQ program is freely available for academic use from Github at  https://github.com/yanxiaodong128/datareconstruction.
\section*{Acknowledgements}
This work was supported in part  by the National Key R$\&$D Program of China (Grant No. 2018YFA0703900) and
the National Natural Science Foundation of China (Grant Num. 11901352).
\section*{Author contributions}
Zengjing Chen conceived and supervised the study. Huaijin Liang supplied the results and method. Wei Wang conducted the experiment,
Xiaodong Yan drafted the parts of abstract, introduction and discussion.
All the
authors read and approved the manuscript.

\section*{Competing interests}
The authors declare no competing interests.

\end{document}